\documentclass[preprint,aps,amsmath,byrevtex,prd,titlepage,superscriptaddress,
tightenlines,
booktabs,
nofootinbib]{revtex4-1}

\usepackage{dcolumn}
\usepackage{amsmath}
\usepackage{amssymb}
\usepackage{bm}
\usepackage{hyperref}
\usepackage{graphicx}
\usepackage{epsfig}
\usepackage{slashed}

\usepackage{mathtools}

\newcolumntype{C}{>{$}c<{$}}

\newcommand{\beq}{\begin{equation}}
\newcommand{\eeq}{\end{equation}}
\newcommand{\eq}[1]{Eq.~(\ref{#1})}

\begin{document}

\title {Decays of Pentaquarks in Hadrocharmonium and Molecular Scenarios}

\author{Michael I.~Eides}
\email[Email address: ]{meides@g.uky.edu}
\affiliation{Department of Physics and Astronomy, University of Kentucky, Lexington, KY 40506, USA}
\affiliation{Petersburg Nuclear Physics Institute, Gatchina, 188300, St.Petersburg, Russia}
\author{Victor Yu.~Petrov}
\email[Email address: ]{Victor.Petrov@thd.pnpi.spb.ru}
\affiliation{Petersburg Nuclear Physics Institute, Gatchina, 188300, St.Petersburg, Russia}


\begin{abstract}

We consider decays of the hidden charm LHCb pentaquarks in the hadrocharmonium and molecular scenarios.  In both pictures the LHCb pentaquarks are essentially nonrelativistic bound states. We develop a semirelativistic framework for calculation of the partial decay widths that allows the final particles to be relativistic. Using this approach we calculate the decay widths in the hadrocharmonium and molecular pictures. Molecular hidden charm pentaquarks are constructed as loosely bound states of charmed and anticharmed hadrons. Calculations show that molecular pentaquarks decay predominantly into states with open charm. Strong suppression of the molecular pentaquark decays into states with hidden charm is qualitatively explained by a relatively large size of the molecular pentaquark. The decay pattern of hadrocharmonium pentaquarks that are interpreted as loosely bound states of excited charmonium $\psi'$ and nucleons is quite different. This time dominate decays into states with hidden charm, but suppression of the decays with charm exchange is weaker than in the respective molecular case. The weaker suppression is explained by a larger binding energy and respectively smaller size of the hadrocharmonium pentaquarks. These results combined with the experimental data on partial decay widths could allow to figure out which of the two theoretical scenarios for pentaquarks (if either) is chosen by nature.

\end{abstract}



\maketitle

\section{Introduction}

Pentaquarks discovered by the LHCb collaboration \cite{LHCb2015,LHCb2016} are the first experimental sighting of exotic baryons. It is probably not by chance that these baryons contain a heavy quark-antiquark pair, with quark masses larger than the scale of strong interactions. Internal structure of the  LHCb pentaquarks remains at this moment unknown. Numerous models of the exotic pentaquarks were proposed in the literature, see, e.g., recent reviews \cite{rflremess2017,als2017,aeapadp2017,slotsdz2018,fkgchugm2018,mkjlrts2018} and references therein.

We will concentrate on the popular molecular and hadrocharmonium scenarios for the LHCb pentaquarks as they were realized in \cite{epp2016,epp2018} (see also \cite{abfs2018}). Neither of these scenarios can be justified on purely theoretical grounds, both are based on  some physically reasonable conjectures about the nature of QCD at low energies. Both in the hadrocharmonium and the molecular pictures pentaquark is assumed to be a nonrelativistic bound state of two hadrons. The main difference between the two models is in the nature of forces that bind constituents into a pentaquark. The idea of the hadrocharmonium picture \cite{sibvol2005,dubvol2008,livol2014} is that almost static heavy quark and antiquark inside an exotic baryon form a small color singlet state -- one of excitations of charmonium. Light valence quarks inside hadrocharmonium also form a color singlet state (nucleon) and occupy a much larger volume. Interaction between an almost static  color singlet heavy quark-antiquark pair and a large color singlet  nucleon is due to the long range color dipole forces and effectively the small static $c\bar c$ pair probes the long wavelength gluon field inside the large light nucleon. Heavy quarkonium interaction with nuclei was considered in \cite{sjbisgt,lms1992}, see also references in \cite{volsh2008}. A QCD motivated potential that depends on the charmonium chromoelectric polarizability and nucleon stress-energy distribution describes charmonium-nucleon interaction, and one can find the spectrum of hidden charm baryons solving the Schr\"odinger equation \cite{epp2016,epp2018}. Literally, the hadrocharmonium picture is justified in the large $N_c$ and heavy quark limit when the mass of the nucleon becomes large and its size remains constant, while the heavy quark-antiquark pair occupies a small volume and is effectively static  \cite{sibvol2005,dubvol2008}.

The molecular scenario of hidden charm pentaquarks initiated in \cite{volok1976} is qualitatively vastly different. In this scenario heavy quark and valence light quark(s) form a color singlet open charm heavy hadron, while the heavy antiquark forms another open charm hadron with the remaining light valence quark(s). These open charm hadrons interact via exchange of light mesons and form a loosely bound pentaquark where the open charm constituent  hadrons and, respectively, heavy quark and antiquark are at rather large distances. The problem with this scenario is that meson exchanges generate attraction at large distances but are too singular at short distances and fail to hold the constituents  far enough to avoid fall to the center. Some kind of hard core should arise and meson exchanges do not provide any effective repulsion at small distances. Therefore the hard core is not under theoretical control while the wave function in the molecular scenario tends to be concentrated there and critically depends on the hard core properties, see, e.g., \cite{epp2018} and references in the reviews \cite{als2017,aeapadp2017,fkgchugm2018}.

Currently both the molecular and hadrocharmonium descriptions of the LHCb pentaquarks are plausible, one cannot choose between them on purely theoretical grounds. Taking into account uncertainty of the theoretical situation, one needs to find experimentally observable signatures  that could help to figure out which of the two scenarios (if any) is realized by nature. In principle, there are many ways to explore  internal structure of hadrons, the most straightforward  approach is just to measure their form factors. Information on the electromagnetic form factors of pentaquarks could immediately resolve the confrontation of the hadrocharmonium and molecular scenarios. However, one cannot expect any experimental data on the form factors of the LHCb pentaquarks any time soon.  The next best option to explore internal structure of pentaquarks is to measure decays widths.

We expect that the dominant contributions to the total width come from two-particle decays. In the hadrocharmonium picture decays with emission of additional pions are strongly suppressed due to small phase volume and pseudogoldstone nature of pions \cite{epp2016}. The constituents of the molecular pentaquark are unstable with respect to decays $D^*\to D+\pi$ and $\Sigma_c\to \Lambda_c+\pi$, and have finite but small widths.  Three-particle decays $P_c(4450)\to \Sigma_c\bar D\pi$ are banned kinematically, $M_{\scriptscriptstyle\Sigma_c}(2455)+M_{\scriptscriptstyle\bar D}(1865)+M_\pi(140)=4460$ MeV$>M_{P_c}(4450)$. Decays $P_c\to \Lambda_c\bar D^*\pi$ are allowed kinematically, $M_{\scriptscriptstyle\Lambda_c}(2286)+M_{\scriptscriptstyle\bar D^*}(1865)+M_\pi(140)=4436$ MeV$<M_{\scriptscriptstyle P_c}(4450)$ but they are suppressed due to a small available phase volume and derivative coupling of pions.

Both in the hadrocharmonium and molecular pictures there are two qualitatively different classes of two-particle pentaquark decay processes. Decays of one kind occur without charm exchange between the constituents and the decay products carry the same charm as the constituents. In decays of the other kind charm is exchanged  and the decay products have charm quantum numbers that do not coincide with the ones of the constituents.

Calculations of the pentaquark decays are impeded by numerous obstacles:  apparent ultraviolet divergences, uncertainty of the cutoff momenta, need to introduce more or less arbitrary form factors, etc. We describe decay processes of nonrelativistic loosely bound pentaquarks  by $t$-channel exchanges between the constituent hadrons\footnote{Processes with the $s$-channel annihilation of heavy $c$-quarks are suppressed due to the Zweig-Okubo-Iizuka rule.}. In transitions without charm exchange interaction is due to the lightest mesons without open charm. In the case when charm of the constituents changes they exchange by the lightest mesons with open charm. A naive expectation is that in each case (hadrocharmonium and molecular pentaquarks) decays without charm exchange dominate and decays with charm exchange are suppressed. This pattern of decays could allow to choose between the hadrocharmonium and molecular pictures of pentaquarks if and when the experimental data for decays will be available.

Let us quantify these expectations.  Notice that to exchange charm the constituents should come very close to each other, at a relative distance $\sim1/m_c$. The probability of this to happen in a nonrelativistic bound state is proportional to $|\psi(0)|^2/m_c^3$, where $\psi(r)$ is the bound state wave function.  But $\psi(0)\sim \kappa^{3/2}$, where $\kappa=\sqrt{2\mu\epsilon}$, $\mu$ is the reduced mass of the system and $\epsilon$ is the binding energy. Then suppression of decays with exchange of charm is described by the factor

\beq
\frac{|\psi(0)|^2}{m_c^3}
=\left(\frac{\mu}{m_c}\right)^\frac{3}{2}\left(\frac{\epsilon}{m_c}\right)^\frac{3}{2}.
\eeq

\noindent
In a hadrocharmonium pentaquark $\mu$ is about the nucleon mass and in a molecular pentaquark $\mu\sim m_c$.
For the $P_c(4450)$ constructed in \cite{epp2016,epp2018} binding energy is $\epsilon\approx 176$ MeV in the hadrocharmonium case, and it is $\epsilon\approx 15$ MeV in the molecular case.  At face value suppression of decays with charm exchange is expected in both pictures and it is stronger in the molecular picture. We will see below that these expectations hold and discuss what happens.

Our principal goal is to find out if measurements of partial widths for decays in the channels with open and hidden charm can help to figure our which of the two scenarios (hadrocharmonium and molecular) of the hidden charm pentaquarks is realized in nature. To this end we develop a semirelativistic approach to calculation of the decays. Let us emphasize that despite bound states both in the hadrocharmonium and the molecular pictures are nonrelativistic, loop momenta are in principle arbitrary and the final decay momentum is sometimes relativistic. In the semirelativistic approach we make a physically reasonable assumption that the intermediate virtual particles in the loop diagrams are always not far from their mass shell what allows to treat them nonrelativistically. On the other hand, our approach allows to treat the exchanged particle as well as the final particles relativistically. Below we consider decays  of the hadrocharmonium and molecular pentaquarks from \cite{epp2018} in this approach. We start with  the basic  features of the semirelativistic approximation that allows one to calculate the pentaquark decays with a reasonable accuracy. We use Feynman diagrams to derive the interaction potentials for different decays, calculate decay widths of hadrocharmonium and molecular pentaquarks\footnote{Decays of pentaquarks in the molecular picture were discussed in the literature earlier, see, e.g., \cite{lsgz2017,sgxz2016,ludong2016,sl2018,kayshs2018} and references therein. To the best of our knowledge decays in the hadrocharmonium picture were never discussed before.}, make  predictions for relative rates  of different decays in each picture and compare the patterns of decays  in hadrocharmonium and molecular scenario.

\section{Semirelativistic Approximation for Pentaquarks Decays\label{genricdec}}

\subsection{Kinematics}

The first task is to derive a practical general formula for calculation of the pentaquark decays. We consider pentaquarks as loosely bound states of two particles with binding energy $\epsilon$  ($M_{\scriptscriptstyle P_c}=M_{\scriptscriptstyle A}+M_{\scriptscriptstyle B}+\epsilon$) much smaller than the reduced mass of the constituents, $|\epsilon|\ll \mu=M_{\scriptscriptstyle A}M_{\scriptscriptstyle B}/(M_{\scriptscriptstyle A}+M_{\scriptscriptstyle B})$. The constituent particles are close to the mass shell and are nonrelativistic, $\epsilon/\mu\sim v^2/c^2$. In the case of the LHCb pentaquark $P_c(4450)$ constructed as a bound state of $\psi'(3686)$ and the nucleon $N(940)$ \cite{epp2016,epp2018} $\mu=749$ MeV, $\epsilon=176$ MeV,  $\epsilon/\mu\sim v^2/c^2\sim 0.23$ and the relativistic correction to the binding  energy is about $v^2/(4c^2)\sim 6$ \%. The accuracy of  the  nonrelativistic approximation for other systems and processes considered below is roughly the same. We will use the nonrelativistic approximation in calculation of widths of loosely bound states ignoring off-masshellness of the constituents. We expect the obtained  results to have error bars about 6-8 \%.

\begin{figure}[h!]
\begin{center}
\includegraphics[width=4 cm]{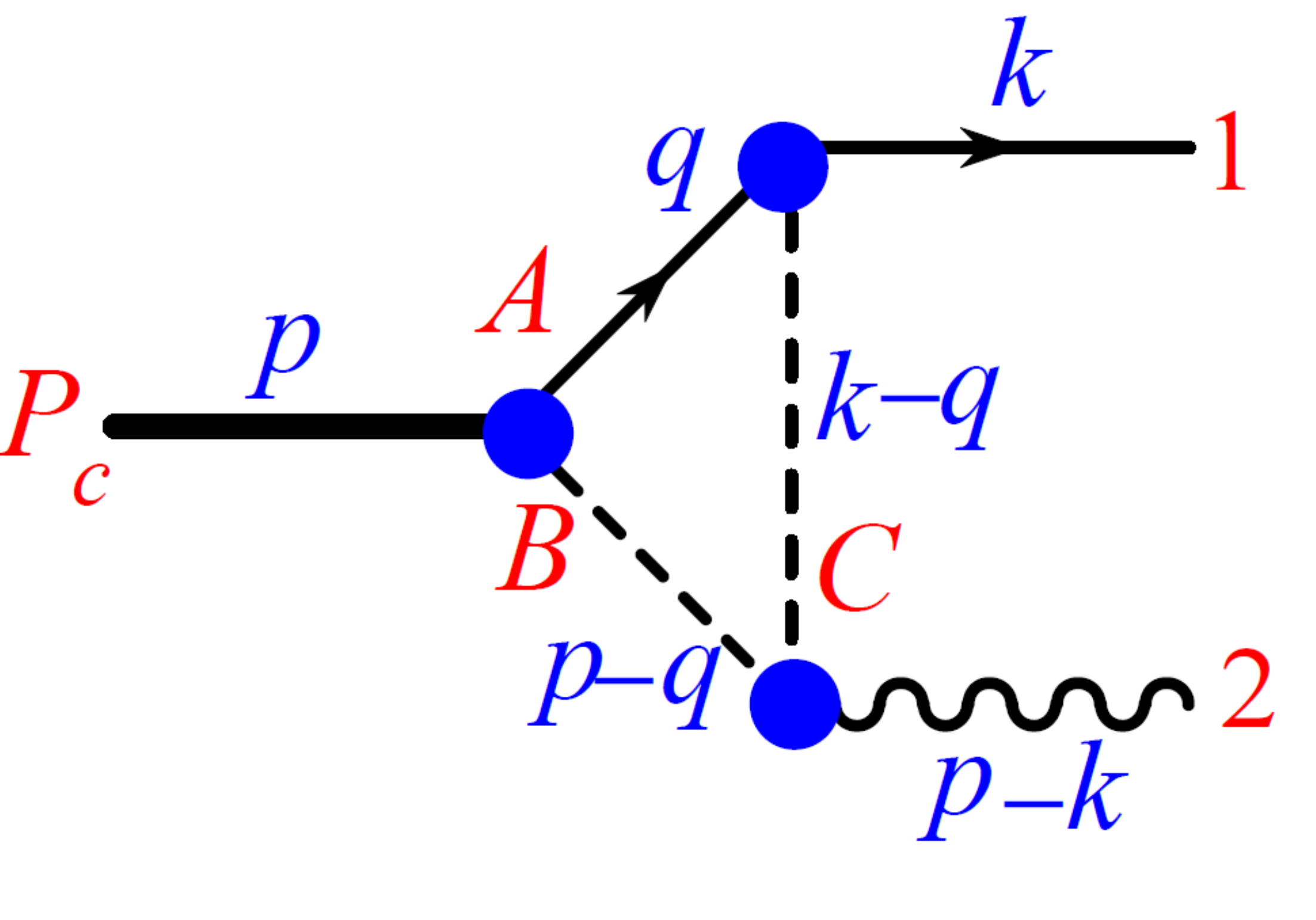}
\end{center}
\caption{Generic diagram for pentaquark decay.\label{figg}}
\end{figure}

Pentaquark decays both in the hadrocharmonium and molecular pictures are due to the diagrams with the $t$-channel exchange  of the type represented in Fig.~\ref{figg}, where $A$ and $B$ are the pentaquark constituents, and $\it 1$ and $\it 2$ are the decay products. To make the discussion more transparent we  temporarily ignore spins of all particles. The final particles  with masses $M_1$ and $M_2$ as well as  the exchanged virtual particle $C$, could have masses significantly smaller than the masses $M_{\scriptscriptstyle A,B}$ of the constituents and are not necessarily nonrelativistic. We need to use relativistic kinematics for these particles. Then the decay width of the pentaquark has the form

\beq \label{decwifhtgen}
\Gamma
=g^2_1g^2_2\frac{k}{4\pi^2}\frac{E_1E_2}{M_{\scriptscriptstyle P_c}}\int d\Omega_k\left|\int d^3re^{-i\bm k\cdot\bm r}V(\bm r,\bm k)\psi(\bm r)\right|^2,
\eeq

\noindent
where $\bm k$ is the three-momentum of the final particle $\it 1$ and we integrate over its directions, $\psi(\bm r)$ is the normalized nonrelativistic wave function of the initial pentaquark (a loosely bound state of particles $A$ and $B$) in its rest frame, and the effective potential  $g_1g_2V(\bm r,\bm k)$ ($g_{1,2}$ are the respective coupling constants) is in the general case a function of the relative coordinate $\bm r$ and the final momentum $\bm k$.  Notice the relativistic energies $E_{1,2}$ in \eq{decwifhtgen} instead of the masses $M_{1,2}$ in the standard nonrelativistic formula. They arise because the final particles could be relatively light and relativistic.

The integral in \eq{decwifhtgen} can be simplified when the bound state wave function $\psi(\bm r)$ is a superposition of terms with different angular momenta $\psi(\bm r)=\sum_lR_l(r)Y_{lm}(\theta,\phi)$ and $V(r)$ is a central potential. In such case  we expand the exponential in spherical harmonics, use their orthogonality and obtain the decay amplitude as a sum of partial waves

\beq \label{sumofpartam}
{\mathcal M}_{if}=\int d^3re^{-i\bm k\cdot\bm r}V(\bm r,\bm k)\psi(\bm r)=
4\pi\sum_l(-i)^lM(l|l)Y_{lm}\left(\frac{\bm k}{k}\right),
\eeq

\noindent
where

\beq \label{matrelll}
M(l|l)=\int_0^\infty r^2 dr R_{l}(r)j_{l}(kr)V(r),
\eeq

\noindent
and $j_{l}(kr)$ is a spherical Bessel function.

The total decay width obtained after integration over angles in this case is

\beq \label{crsectinbessels}
\Gamma =g_1^2g_2^2\frac{4kE_1E_2}{M_{\scriptscriptstyle P_c}}\sum_l|M(l|l)|^2.
\eeq

\noindent
In the calculations below the interaction potential is often a tensor, so the matrix elements  similar to $M(l|)$ are nondiagonal in $l$, in other words orbital momentum changes in decays. The total angular momentum with account for spins is of course conserved.

The effective potential  $V(\bm r,\bm k)$

\beq \label{fourefpot}
V(\bm r,\bm k)=\int\frac{d^3q}{(2\pi)^3}e^{i\bm q\cdot\bm r}V(\bm q,\bm k)
\eeq

\noindent
can be calculated in terms of the relativistic scattering amplitude ${\mathcal A}_{A+B\to 1+2}(\bm q,\bm k)$ with the nonrelativistic initial particles

\beq \label{potintermamp}
g_1g_2V(\bm q,\bm k)=-\frac{{\mathcal A}_{\scriptscriptstyle A+B\to 1+2}(\bm q,\bm k)}{\sqrt{2M_{\scriptscriptstyle A}}\sqrt{2M_{\scriptscriptstyle B}}\sqrt{2E_1}\sqrt{2E_2}}.
\eeq

\noindent
The square roots in this relationship convert the relativistically normalized  scattering amplitude to the normalization used in nonrelativistic quantum mechanics. It is convenient to rescale the potential so that it coincides with the amplitude ${\mathcal A}_{A+B\to 1+2}(\bm q,\bm k)$

\beq \label{replcesqr}
V(\bm q,\bm k)\to \frac{V(\bm q,\bm k)}{\sqrt{2M_{\scriptscriptstyle A}}\sqrt{2M_{\scriptscriptstyle B}}\sqrt{2E_1}\sqrt{2E_2}}.
\eeq

\noindent
Then the total width in \eq{crsectinbessels} acquires the form

\beq \label{gentrtotw}
\Gamma =g_1^2g_2^2\frac{4kE_1E_2}{M_{\scriptscriptstyle P_c}}\frac{\sum_l|M(l|l)|^2}{2M_{\scriptscriptstyle A}2M_{\scriptscriptstyle B}2E_12E_2}.
\eeq

\noindent
Below we will use a natural generalization of this formula for particles with spin.

Our strategy is to use the standard Feynman rules with free initial and final particles to calculate the scattering amplitudes with the nonrelativistic initial particles. Then we convert the scattering amplitudes into effective potentials $V(\bm r,\bm k)$, expand the integrand in  \eq{decwifhtgen} in spherical harmonics (with account for spin, if necessary), calculate the angular integrals analytically and finish with computing the remaining radial integrals numerically, using the wave functions obtained in \cite{epp2018}.

Let us illustrate the logic of calculations still assuming that all particles in  Fig.~\ref{figg} are scalars. In this case the rescaled potential is just

\beq
V(\bm k,\bm q)=\frac{1}{M_{\scriptscriptstyle C}^2-(k-q)^2}.
\eeq

\noindent
All external momenta are on mass shell and

\beq \label{defmstar}
M_{\scriptscriptstyle C}^2-(k-q)^2=\left[M_{\scriptscriptstyle C}^2-\left(M_{\scriptscriptstyle A}-\sqrt{M_1^2+\bm k^2}\right)^2\right]+(\bm k-\bm q)^2
\equiv M_*^2(C)+(\bm k-\bm q)^2,
\eeq

\noindent
and

\beq
V(\bm k,\bm q)=\frac{1}{M_*(C)^2+(\bm k-\bm q)^2}.
\eeq

\noindent
In this simple case the potential is a function only of $(\bm k -\bm q)^2$ and its Fourier transform is just the Yukawa potential. Notice that its radius is determined not by the mass of the exchanged particle $M_{\scriptscriptstyle C}$ but by the effective mass $M_*(C)=\sqrt{M_{\scriptscriptstyle C}^2-\left(M_{\scriptscriptstyle A}-\sqrt{M_1^2+\bm k^2}\right)^2}$.

\subsection{Tensor, Spin, and Isospin Structure of Decay Potentials}

In the nonrelativistic approximation one-pion exchange in Fig.~\ref{figg}  generates a relatively long-range effective potential between $\Sigma_c$ and $\bar D^*$  that was used in \cite{epp2018} in discussion of the molecular pentaquark

\beq \label{onepionpot}
V(\bm q)=-4\frac{g^{\scriptscriptstyle A}_{\scriptscriptstyle \Sigma_c}g^{\scriptscriptstyle A}_{\scriptscriptstyle D^*}}{F_\pi^2}(\bm t_1\cdot\bm t_2)\frac{(\bm s^{(1)}\cdot\bm q)(\bm s^{(2)}\cdot\bm q)}{m_\pi^2+\bm q^2},
\eeq

\noindent
where $g^{\scriptscriptstyle A}_{\scriptscriptstyle \Sigma_c}$ and $g^{\scriptscriptstyle A}_{\scriptscriptstyle D^*}$ are the axial charges of $\Sigma_c$ and $D^*$, respectively, and matrix elements of the spin and isospin operators $\bm t_i$ and $\bm S_i$ should be calculated between the state vectors of the respective particles. In coordinate space the momentum-dependent factor turns into a superposition of a central and tensor potentials (we temporarily omit the coupling constants)

\beq \label{tensorpotcoor}
W_{ij}(\bm r)=4\int \frac{d^3q}{(2\pi)^3}
\frac{q_iq_j}{m_\pi^2+\bm q^2}e^{i\bm q\cdot\bm r}
=V_c(r)\delta_{ij}+(3n_in_j-\delta_{ij})V_t(r),
\eeq

\noindent
where $n_i=r_i/r$ and

\beq \label{scaltenpot}
V_c(r)=\frac{m^2e^{-mr}}{3\pi r},
\qquad
V_t(r)=\left[3+3mr+(mr)^2\right]\frac{e^{-mr}}{3\pi r^3}.
\eeq

\noindent
There is also an additional term proportional to $\delta(\bm r)$ on the right hand side in \eq{tensorpotcoor}. We omit it as unphysical in calculations of the bound state energies, because it arises from the distances  where the one-pion exchange makes no sense due to finite sizes of all particles, see \cite{epp2018} for details. The spin and isospin matrices in \eq{onepionpot} act in the space of spin and isospin states of the constituents. In \cite{epp2018} we used the potentials in \eq{onepionpot} and \eq{scaltenpot}  together with the similar  potentials that arise from $\sigma$, $\rho$, $\omega$ and $\eta$ exchanges to construct a loosely bound pentaquark state $P_c(4450)$. All potentials were regularized at small distances about 0.15 fm, for details of the regularization see  Eq.(31,32) in \cite{epp2018}.

Decays of molecular pentaquarks  without charm exchange can go via exchanges by a pion and other light mesons. We expect that the one-pion contribution, without account for exchanges by other mesons, gives a reasonable estimate of  decay widths. Unlike the case of the binding potential, one-particle exchange decay amplitudes  describe transitions from one pair of particles to another.  After calculations pion exchange reduces to the potentials of the same type as in   \eq{tensorpotcoor} and \eq{scaltenpot}, the only differences are that we use the nondiagonal axial charges (see also \cite{hpkmw2017}), and substitute $m_\pi\to m_*(\pi)$ and $\bm q^2\to (\bm k-\bm q)^2$, compare \eq{defmstar}. Decays of the molecular and hadrocharmonium  pentaquarks with exchange of charm go via $D$-meson and other heavy hadron exchanges. The respective effective potentials do not coincide with the ones in \eq{tensorpotcoor} and \eq{scaltenpot}, but still depend on spin, isospin and orbital momenta. This allows us to give a universal description of the strategy of further calculations. Consider, for example, a molecular pentaquark decay. The bound state wave function  of the molecular pentaquark \cite{epp2018} is a superpositions of the states  $|l=0,S=3/2\rangle$,  $|l=2,S=1/2\rangle$, and $|l=2,S=3/2\rangle$, where $l$ is the orbital momentum and $S$ is the total spin of the pentaquark. Each of the components of the molecular $\Sigma_c\bar D^*$ wave function is in its turn a superposition of one-particle spin-isospin states of the constituents.  In terms of these spin-isospin states of the constituents the $\Sigma_c\bar D^*$ the wave function of the pentaquark in the state $|j={3}/{2},j_3;t={1}/{2},t_3\rangle$ has the form

\beq \label{intitwavef}
\Psi^{\scriptscriptstyle\frac{3}{2},j_3;\frac{1}{2},t_3}(\bm r)=\sum
C^{\frac{3}{2}j_3}_{SS_3,lm}C^{SS_3}_{\frac{1}{2}s^{(1)}_3,1s^{(2)}_3}
C^{\frac{1}{2}t_3}_{1t^{(1)}_3,\frac{1}{2}t^{(2)}_3}
R_{lS}(r)Y_{lm}(\bm n)\Sigma_{s^{(1)}_3t^{(1)}_3}\bar{D}^*_{s^{(2)}_3t^{(2)}_3},
\eeq

\noindent
where $\Sigma_{s^{(1)}_3t^{(1)}_3}$ and $\bar{D}^*_{s^{(2)}_3t^{(2)}_3}$ are normalized to unity spin-isospin states  of $\Sigma_c$ and $\bar D^*$ with the spin projection $s^{(i)}_3$ and the isospin  projection $t^{(i)}_3$, $j_3,t_3$ are the third components of the pentaquark spin and isospin, $Y_{lm}(\bm n)$ are spherical harmonics, $C^{\frac{3}{2}j_3}_{SS_3,lm}$, $C^{SS_3}_{\frac{1}{2}s^{(1)}_3,1s^{(2)}_3}$,
$C^{\frac{1}{2}t_3}_{1t^{(1)}_3,\frac{1}{2}t^{(2)}_3}$ are the Clebsch-Gordan coefficients, and $R_{lS}(r)$ are the radial wave functions in the states $|l,S\rangle$. Summation runs over spin and isospin projections of the constituents  and  includes also summation over three available $l,S$ combinations.

We consider a one-particle exchange scattering amplitude as an operator that acts on the initial wave function in \eq{intitwavef} and transforms it in a superpositions of products of spin-isospin one-particle states of the final particles with the coefficients that are  coordinate wave functions of their relative motion. Like in \eq{intitwavef} these coordinate wave functions are themselves superpositions of products of radial wave functions and spherical harmonics. The final orbital momenta arise automatically by addition of orbital momenta of the initial wave function and of the interaction potential and do not coincide with the initial orbital momenta, only the total angular momentum is conserved in the general case. Next we project this wave function on the final plane wave, compare \eq{sumofpartam}. We obtain a superposition of matrix  elements of the potential $M(l,S|L)$ (compare \eq{matrelll}), with the coefficients that are spin-isospin wave functions of the final particles.  Unlike the expression in \eq{matrelll} the radial wave function $R_{lS}(r)$ carries now a second index $S$ because it depends on the total spin of the bound state. In addition the final angular momentum $L$ in the integral for $M(l,S|L)$ does not necessarily coincide  with the initial angular momentum $l$ since the potential is in the general case a coordinate space (as well as spin and isospin) tensor. These matrix elements $M(l,S|L)$ are decay amplitudes of the initial state $|l,S\rangle$ into a final state with the total orbital momentum $L$ and spin-isospin quantum numbers of the coefficients.

To calculate the decay width in any channel we apply the operator arising from the respective one-particle exchange amplitude to the wave function \eq{intitwavef} of the pentaquark with fixed quantum numbers. Then we obtain the decay amplitude as a superposition of matrix elements $M(l,S|L)$, square it, calculate the integrals over directions of the final momentum $\bm k$   and thus obtain the decay width. We will fill some technical gaps in this schematical discussion considering the decays below.

\section{Decays of Molecular Pentaquarks}

Let us recall the principal features of the molecular pentaquark scenario considered in \cite{epp2018}.
Exotic pentaquarks in this picture are loosely bound states of  hadrons with open charm located at rather large distances. One could expect that the interaction of the constituent hadrons in this case would be dominated by the long-range one-pion exchange and the pentaquark would resemble the deuteron, see, e.g.,  \cite{nat1991}. We considered this binding mechanism in \cite{epp2018} and came to the conclusion that the effective distances are not large enough to neglect exchanges by other light mesons, besides pions. The pion exchange in \cite{epp2018} was regularized to get rid of its unphysical too singular behavior at small distances, and exchanges by $\sigma$, $\rho$, $\omega$ and $\eta$ were also taken into account. Then we constructed the pentaquark $P_c(4450)$ as a loosely bound state of  $\Sigma_c(2455)$ ($I(J^P)=1({1}/{2}^+)$) and $\bar{D}^*(2010)$ ($I(J^P)=1/2(1^-)$) with the binding energy only 15 MeV and spin-parity $({3}/{2})^-$. This pentaquark arises when the regularization parameter $\Lambda =1300$ MeV, with the root mean square radius $1.46$ fm and $D$-wave squared fraction about 4\%, see \cite{epp2018} for more details. An attempt to use the potential with the same parameters in order to construct $P_c(4380)$ as a loosely bound state of ${\Sigma}^*_c(2520)$ ($I(J^P)=1({3}/{2}^+)$) and $\bar{D}(1870)$ ($I(J^P)=1/2(0^-)$) with the binding energy 10 MeV was not successful. The main reason is that the would be constituents ${\Sigma}^*_c$ and $\bar{D}$ do not interact via one-pion exchange since the three-pseudoscalar vertex $\pi DD$ is banned by parity, and exchanges by the other light mesons cannot provide the necessary binding. Therefore, if we insist that the LHCb $P_c(4380)$ pentaquark should be a loosely bound molecular state with a tiny binding energy its nature in this picture remains an open question.

Small binding energy and large size of the molecular pentaquark $P_c(4450)$ imply that the constituent hadrons are non-relativistic and this bound state can be described in the potential approach. We constructed such molecular pentaquark in \cite{epp2018}. Let us consider its decays due to one-particle exchanges.

\subsection{Decays into States with Open Charm }

There are four open channels for the $P_c(4450)$ pentaquark decays  into states with open charm, see Table~\ref{pentmoldecop}.  In the case of the molecular pentaquark there is no charm exchange in these decays and they can go via one-pion exchanges. As mentioned above, exchanges by heavier mesons are also allowed but we will account only for the contribution of the pion exchange.

\subsubsection{$P_c\to \Lambda_c+\bar D$ Decay}

\begin{figure}[h!]
\begin{center}
\includegraphics[width=4 cm]{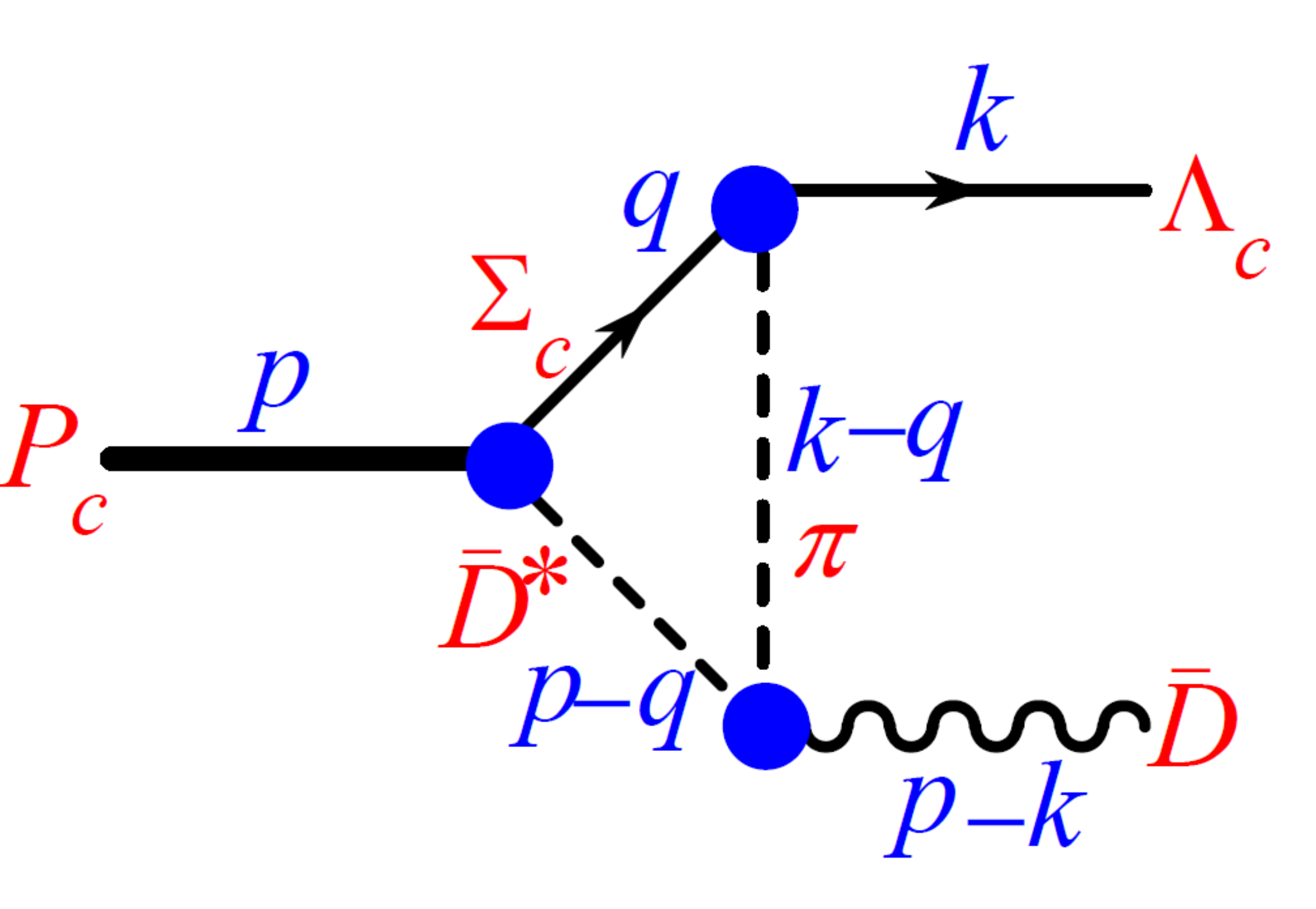}
\end{center}
\caption{Decay of molecular pentaquark $P_c(4450)$ into open charm states $\bar{D}+\Lambda_c$}
\label{moldeclambd}
\end{figure}

We start with the channel $P_c\to \Lambda_c+\bar{D}$. The initial pentaquark has spin-parity $3/2^-$ and isospin $1/2$, the final $\Lambda_c$ carries spin-parity $1/2^+$ and zero isospin, and the final $\bar D$ is a pseudoscalar with isospin $1/2$. The product of the internal parities of $\Lambda_c$ and $\bar D$ is negative, so the final state in the decay $P_c(4450)\to \Lambda_c+\bar D$ can have only even angular momenta. The final state with $L=0$ is banned by the angular momentum conservation, so the lowest allowed final orbital momentum is $L=2$. The final decay momentum is $k\approx 798$ MeV, and both final particles are nonrelativistic with a reasonable accuracy,  $E_\Lambda\approx2421$ MeV and $(E_\Lambda-M_\Lambda)/M_\Lambda\approx0.059$, and $E_{\bar D}\approx2029$ MeV and $(E_{\bar D}-M_{\bar D})/M_{\bar D}\approx0.087$.

\begin{figure}[h!]
\begin{center}
\includegraphics[width=3 cm]{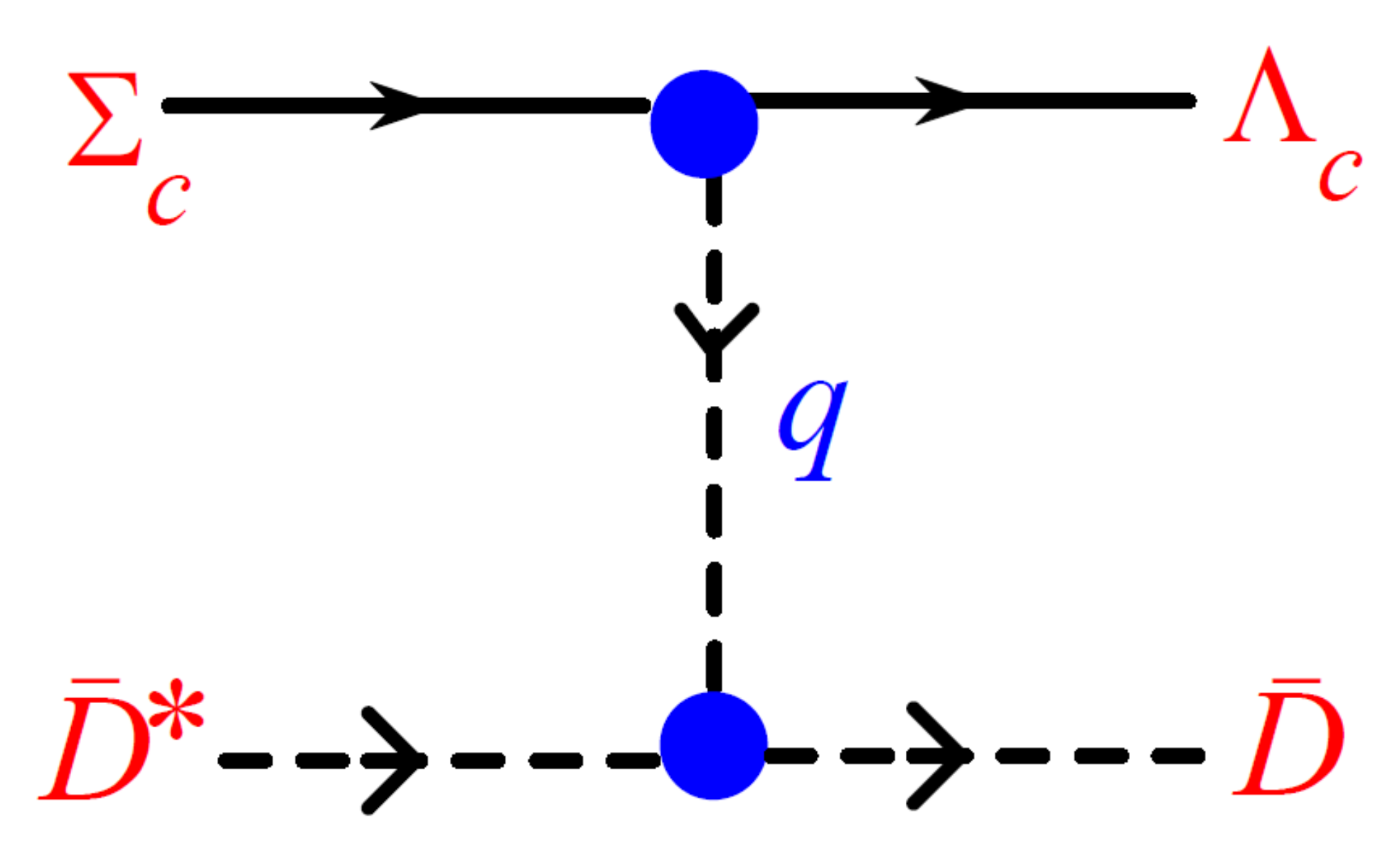}
\end{center}
\caption{Amplitude $\Sigma_c+\bar D^*\to \Lambda_c+\bar D$}
\label{sigmcbdslambd}
\end{figure}

This decay is described by the diagram in Fig.~\ref{moldeclambd}. First we calculate the relativistic scattering amplitude in Fig.~\ref{sigmcbdslambd}

\beq
{\mathcal A}(\bm q,\bm k)
=g_{\scriptscriptstyle\pi\Sigma_c\Lambda_c}g_{\scriptscriptstyle\pi D D^*} \bar\Lambda_c(\bm k)\gamma^5 \frac{(k-q)_\nu}{m_\pi^2-(k-q)^2}
\Sigma_c^a\bar D^\dagger\tau_a D^{*\nu}(\bm q),
\eeq

\noindent
where $\bar D^{*\nu}(\bm q)$ is a four-vector isospinor, $\bar D$ is an isospinor, $\Sigma^a_c$ is a spinor  isovector, and $\Lambda_c(\bm k)$ is a spinor. The coupling constants and interaction Lagrangians can be found in Table~\ref{pionlgr} and are discussed in Appendix~\ref{pionintrco}.

In the nonrelativistic approximation the denominator of the propagator reduces to $m_*^2(\pi)+(\bm k-\bm q)^2$, and the interaction radius is determined by $m_*(\pi)=\{m_\pi^2-[(M_{\Lambda_c}^2+\bm k^2)^\frac{1}{2}-M_{\Sigma_c}]^2\}^\frac{1}{2}=136$ MeV. Using this approximation  for the the initial and final particles and omitting the coupling constants and certain square roots of masses (to be restored in the final expression for the decay width, compare \eq{replcesqr}) we obtain the interaction potential that acts as an operator on the initial pentaquark wave function in \eq{intitwavef}

\beq  \label{trnasopsdsld}
\left(\Lambda_c^\dagger\sigma^i\Sigma_c^a\right)W_{ik}(\bm k-\bm q)
\left(\bar D^\dagger\tau^a\bar D^{*k}\right),
\eeq

\noindent
or in coordinate space

\beq \label{potenasopdeld}
\left(\Lambda_c^\dagger\sigma^i\Sigma_c^a\right)W_{ik}(\bm r)
\left(\bar D^\dagger\tau^a\bar D^{*k}\right),
\eeq

\noindent
where $W_{ik}(\bm r)$ is defined in \eq{tensorpotcoor} (now with $m\to m_*(\pi)$) and $\bar D,\bar D^{*k},\Sigma_c^a,\Lambda_c$ are nonrelativistic spin-isospin states similar to the ones in \eq{intitwavef}.

It is convenient to represent $W_{ik}$ in terms of spherical harmonics\footnote{We use conventions for spherical harmonics from \cite{ll1991}, in particular

\beq
Y_{00}=\frac{1}{\sqrt{4\pi}},\quad Y_{20}=\sqrt{\frac{5}{16\pi}}(1-3n_3^2),\quad Y_{2,\pm1}=\pm\sqrt{\frac{15}{8\pi}}n_3(n_1\pm in_2),\quad Y_{2,\pm2}=-\sqrt{\frac{15}{32\pi}}(n_1\pm in_2)^2.
\eeq
}

\beq
W_{m_1m_2}(\bm r) =
V_c(r)(-1)^{1-m_1}\delta_{m_1,-m_2} -V_t(r)\sqrt{\frac{24\pi}{5}}C^{1,m_1+m_2}_{1m_1,1m_2} Y_{2,-m_1-m_2},
\eeq

\noindent
where $V_c(r)$ and $V_t(t)$ are the regularized potentials in \eq{scaltenpot}, see discussion of the regularization below \eq{scaltenpot} and in \cite{epp2018}.

The transition operator in \eq{potenasopdeld} should be applied to the initial wave function of  the molecular pentaquark. We choose the initial pentaquark state with $j_3=3/2$  and $t_3=1/2$. The interaction operator in  \eq{potenasopdeld} transforms it into the final wave function. After projection on the final plane wave and spatial integration we obtain the decay amplitude

\beq \label{fintwafrlamd}
\begin{split}
{\mathcal M}_{i\to f}&=\frac{3}{\sqrt{5}}\left[{M_c\left(2,\frac{1}{2}\biggl|2\right)} +{M_t\left(0,\frac{3}{2}\biggl|2\right)}
-{M_t\left(2,\frac{3}{2}\biggl|2\right)}\right] Y_{21}(\bm{n})\bar{D}^{0\dagger}\Lambda_c^\dagger\left[\frac{1}{2}\right] \\
&-\frac{6}{\sqrt{5}}\left[{M_c\left(2,\frac{1}{2}\biggl|2\right)} +{M_t\left(0,\frac{3}{2}\biggl|2\right)}
-{M_t\left(2,\frac{3}{2}\biggl|2\right)}\right] Y_{22}(\bm{n})\bar{D}^{0\dagger}\Lambda_c^\dagger\left[-\frac{1}{2}\right],
\end{split}
\eeq

\noindent
where $\Lambda_c^\dagger[\pm{1}/{2}]$ is the final $\Lambda_c$ with spin up or down, $\bm n=\bm k/|\bm k|$, and
$M_{c,t}(l,S|L)$ are radial matrix elements of the potentials $V_{c,t}$ between the initial pentaquark state $|l,S\rangle$ and the final two-particle  state with the orbital momentum $L=2$ similar to the ones in \eq{matrelll}. We see that interaction in \eq{potenasopdeld} generates only the transitions to the final states in $D$-wave. Next we calculate module  square of the transition matrix element in \eq{fintwafrlamd}, integrate over the directions of the final momentum, and sum over all allowed final states

\beq \label{summelsqld}
{\int\mathllap{\sum}}_f \:
|{\mathcal M}_{i\to f}|^2 = 9 \left|M_c\left(2,\frac{1}{2}\biggl|2\right)+M_t\left(0,\frac{3}{2}\biggl|2\right)
-M_t\left(2,\frac{3}{2}\biggl|2\right)\right|^2.
\eeq

\noindent
The decay width is calculated with a natural generalization of \eq{gentrtotw}

\beq \label{generalgamma}
\Gamma = g_1^2g_2^2\frac{4kE_1 E_2}{M_{P_c}}\frac{{\int\mathllap{\sum}}_f{|\mathcal
M}_{i\to f}|^2}{(2M_1)(2M_2)(2M_{\scriptscriptstyle A})(2M_{\scriptscriptstyle B})},
\eeq

\noindent
where we plug in $g_1=g_{\scriptscriptstyle\pi\Sigma_c\Lambda_c}$, $g_2=g_{\scriptscriptstyle\pi D D^*}$, $M_{\scriptscriptstyle A}=M_{\scriptscriptstyle\Sigma_c}$, $M_{\scriptscriptstyle B}=M_{\scriptscriptstyle D^*}$, $M_1=M_{\scriptscriptstyle\Lambda_c}$, $M_2=M_{\scriptscriptstyle D}$, $E_1=\sqrt{M_{\scriptscriptstyle\Lambda_c}^2+\bm k^2}$, $E_2=\sqrt{M_{\scriptscriptstyle D}^2+\bm k^2}$, and sum of matrix elements squared from \eq{summelsqld}. We use \eq{generalgamma} for calculations of all decay widths below.

After numerical calculations we obtain $\Gamma(P_c\to \Lambda_c+\bar D)=6.8$ MeV.

\subsubsection{Other Open Charm Decays of Molecular Pentaquark }

Calculation of other three decays of the molecular pentaquark into states with an open charm

\beq
P_c\to \Sigma_c+\bar D,\qquad P_c\to \Lambda_c+\bar D^*,\qquad P_c\to \Sigma^*_c+\bar D,
\eeq

\noindent
is similar to the calculations above. All these decays go via the pion exchange,  the final decay momenta are even smaller than in the decay $P_c\to\Lambda_c+\bar D$, see Table~\ref{pentmoldecop}, and the decay products are nonrelativistic.


Decay $P_c\to \Sigma_c+\bar D$ requires almost no new calculations.
Spin-parity of $\Sigma_c(2455)$ are the same as spin-parity of $\Lambda_c$ and like in the previous decay $L=2$ is the lowest allowed partial wave. The final momentum is $k\approx 529$ MeV, and  the final particles are again essentially nonrelativistic. Kinetic energy of the $D$-meson is about 4\% of its mass,  and kinetic energy of $\Sigma_c$ is about 2\% of its mass.

\begin{figure}[h!]
\begin{center}
\includegraphics[width=4 cm]{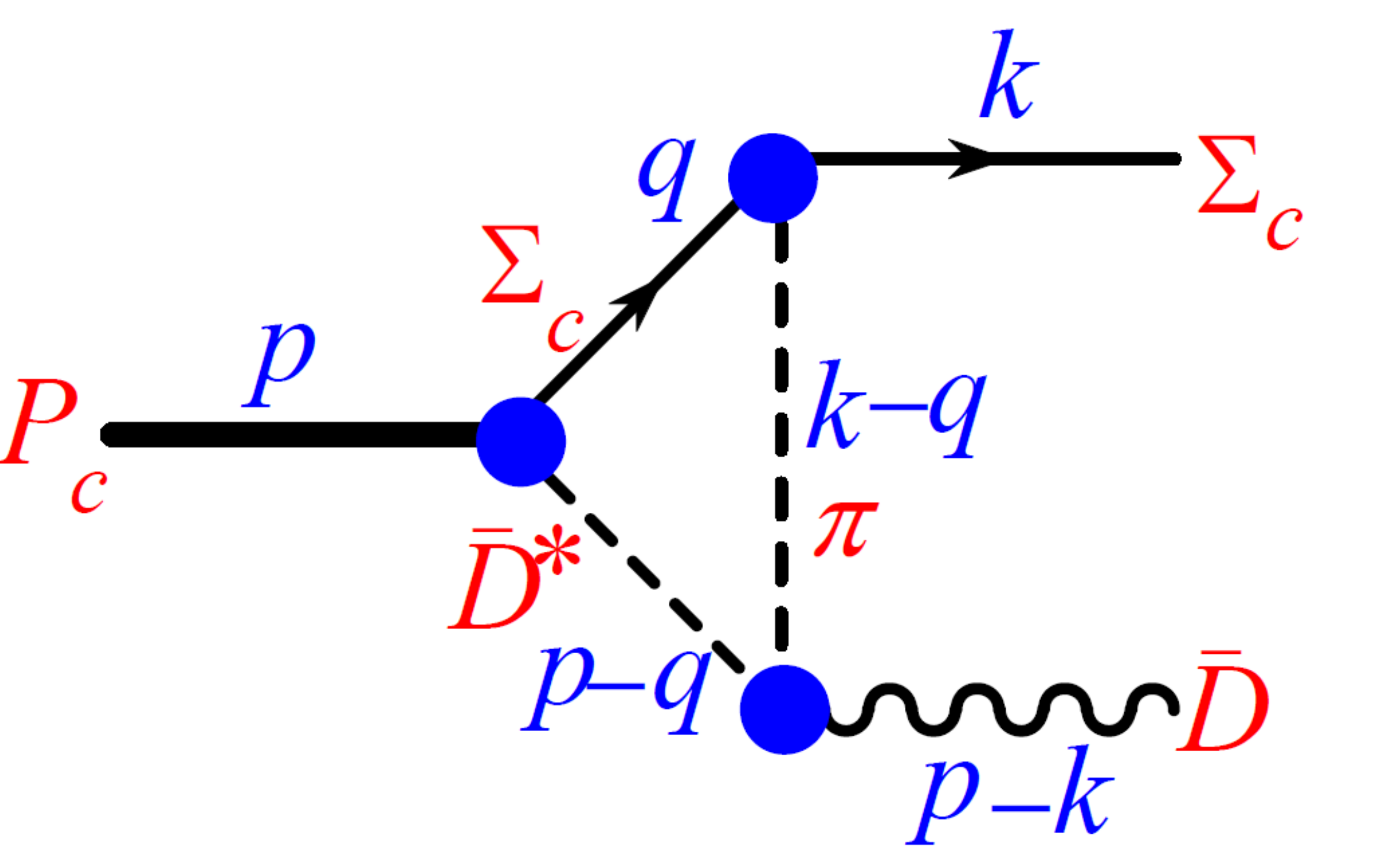}
\end{center}
\caption{Decay of  molecular pentaquark $P_c(4450)$ into open charm states $\bar{D}+\Sigma_c$}
\label{fig6}
\end{figure}

The $P_c\to \Sigma_c+\bar D$ decay amplitude in Fig.~\ref{fig6} can be obtained from the decay amplitude $P_c\to \Lambda_c+\bar D$ in Fig.~\ref{moldeclambd}. Only the isotopic structure of the $\pi\Sigma_c\Sigma_c$  vertex is different from the isotopic structure of the $\pi\Lambda_c\Sigma_c$ vertex, see the respective interaction Lagrangians in Table~\ref{pionlgr}. The isotopic factor factorizes in the decays amplitudes and the decay width of  $P_c\to \Sigma_c+\bar D$ is equal to the decay width of $P_c\to \Lambda_c+\bar D$ times the ratio  of the respective isotopic factors squared.

The isospinor isotopic factor in the molecular pentaquark wave function is $\Psi^{iso}_\alpha=(1/\sqrt{3})\Sigma^a_c(\tau^a)_{\alpha\beta}\bar D^*_\beta$.  In the case of $P_c\to\Lambda_c+\bar D$ decay we apply to this wave function the isotopic factor $\tau^a$ in the transition operator in \eq{potenasopdeld} and obtain the final isotopic function

\beq
\Psi_{fin}^{iso,\alpha}(\bar{D}+\Lambda_c) = \frac{1}{\sqrt{3}}(\tau_a\tau_a)^\alpha_\beta
\bar{D}^{\beta}\Lambda_c = \sqrt{3}\delta^\alpha_\beta \bar{D}^{\beta}\Lambda_c.
\eeq

\noindent
In the case of the $P_c\to\Sigma_c+\bar D$ decay the isotopic factor in the transition operator in  the diagram in Fig.~\ref{fig6} is $\tau_a\epsilon_{abc}$ and then the final isotopic wave function is

\beq
\Psi_{fin}^{iso,\alpha}(\bar{D}+\Sigma_c) =
\frac{1}{\sqrt{3}}(\tau^b\tau^c)^\alpha_\beta\varepsilon_{abc} \bar{D}^{\beta}\Sigma^a_c
=\frac{2i}{\sqrt{3}}(\tau^a)^\alpha_\beta\bar{D}^{\beta}\Sigma^a_c.
\eeq

\noindent
Squaring the isotopic factors in the scattering amplitudes and summing over all allowed final isotopic states we obtain the isotopic factor contributions to the decay width in both cases

\beq
\Phi^{iso}(P_c\to\Lambda_c+\bar D)=3,\qquad \Phi^{iso}(P_c\to\Sigma_c+\bar D)=\frac{4}{3} (\tau^a\tau^a)^\alpha_{\alpha}=4.
\eeq

\noindent
Spin and orbital structure of the matrix elements is identical for both decays. Hence, the sum of matrix elements squared for the decay $P_c\to\Sigma_c+\bar D$ is $4/3$ times larger than the sum of matrix elements squared for the decay $P_c\to\Lambda_c+\bar D$, and (compare \eq{summelsqld})

\beq
{\int\mathllap{\sum}}_f \:
|{\mathcal M}_{i\to f}|^2
= 12 \left|M_c\left(2,\frac{1}{2}\biggl|2\right)+M_t\left(0,\frac{3}{2}\biggl|2\right)
-M_t\left(2,\frac{3}{2}\biggl|2\right)\right|^2
\eeq

\noindent
for $P_c\to\Sigma_c+\bar D$.

Calculating the width according to \eq{generalgamma} we obtain $\Gamma(P_c\to\Sigma_c+\bar D)=1.4$ MeV.

\begin{figure}[h!]
\begin{center}
\includegraphics[width=4 cm]{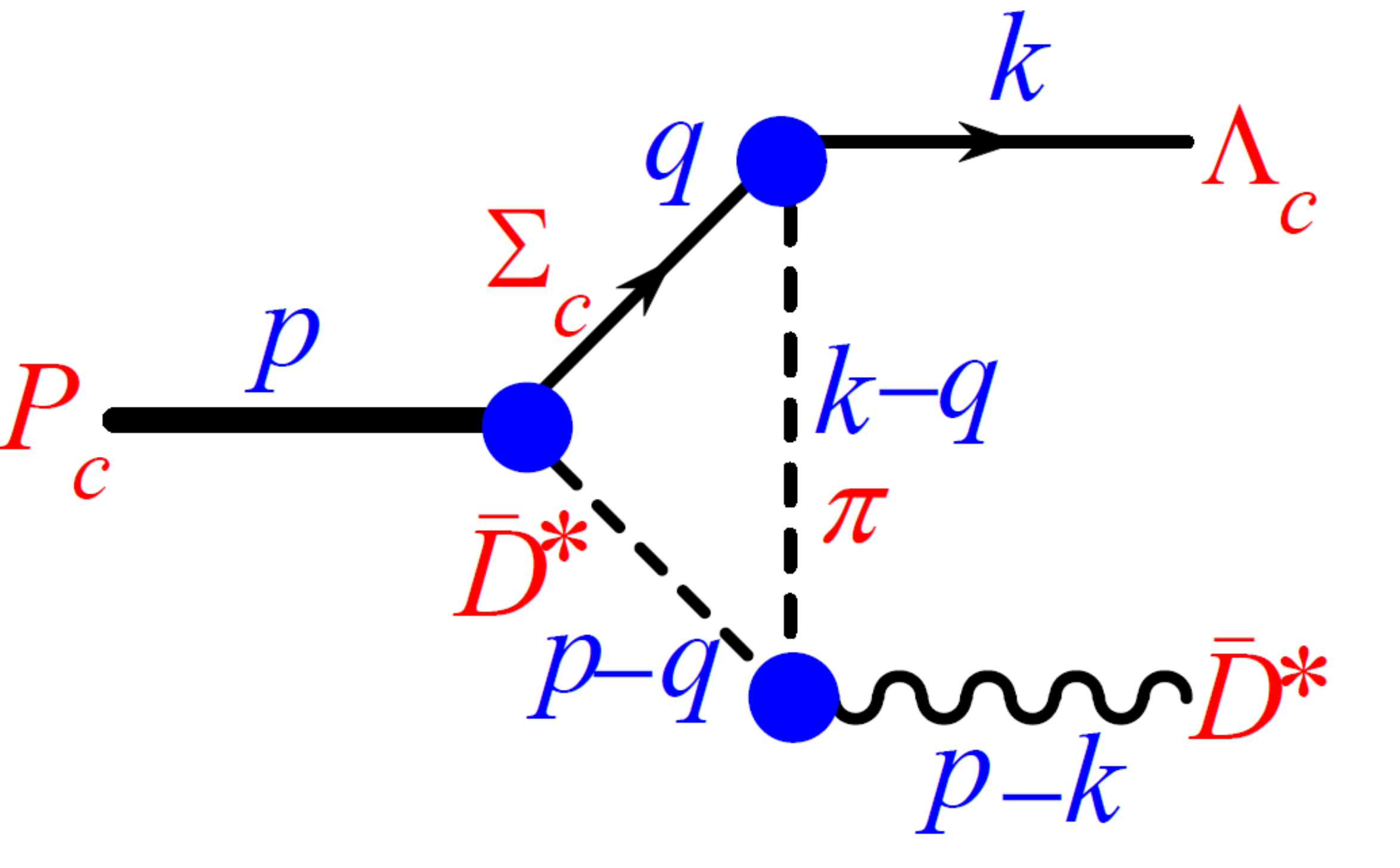}
\end{center}
\caption{Decay of molecular pentaquark $P_c(4450)$ into open charm states $\bar{D}^*+\Lambda_c$}
\label{fig7}
\end{figure}

The $P_c\to \Lambda_c+\bar D^*$ decay  goes via the one-pion exchange diagram  in Fig.~\ref{fig7}. The $D^*D^*\pi$   interaction Lagrangian and coupling constant are in Table~\ref{pionlgr}. Let us notice that both interaction constants in this decay are found from the experimental data on decays, see discussion in Appendix~\ref{pionintrco}.

We go through by now the standard steps and obtain a rather cumbersome sum of matrix elements squared for this decay

\beq
\begin{split}
{\int\mathllap{\sum}}_f \:
&|{\mathcal M}_{i\to f}|^2\\
& = \frac{3}{5}\left|
M_c\left(2,\frac{3}{2}\biggl|2\right)+2M_t\left(0,\frac{3}{2}\biggl|2\right)
-M_t\left(2,\frac{1}{2}\biggl|2\right)\right|^2\\
&+3\left| M_c\left(0,\frac{3}{2}\biggl|0\right)+M_t\left(2,\frac{1}{2}\biggl|0\right)
+2M_t\left(2,\frac{3}{2}\biggl|0\right)\right|^2\\
& +
\frac{1}{5}\left|2M_c\left(2,\frac{1}{2}\biggl|2\right)+2M_c\left(2,\frac{3}{2}\biggl|2\right)
 +3M_t\left(0,\frac{3}{2}\biggl|2\right)-2M_t\left(2,\frac{1}{2}\biggl|2\right)+M_t\left(2,\frac{3}{2}\biggl|2\right) \right|^2\\
& +
\frac{6}{5}\left|2M_c\left(2,\frac{1}{2}\biggl|2\right)-M_c\left(2,\frac{3}{2}\biggl|2\right)
-3M_t\left(0,\frac{3}{2}\biggl|2\right)+M_t\left(2,\frac{1}{2}\biggl|2\right)+M_t\left(2,\frac{3}{2}\biggl|2\right) \right|^2\\
& +
\frac{2}{5}\left|4M_c\left(2,\frac{1}{2}\biggl|2\right)+M_c\left(2,\frac{3}{2}\biggl|2\right)
-M_t\left(2,\frac{1}{2}\biggl|2\right)+2M_t\left(2,\frac{3}{2}\biggl|2\right)\right|^2.
\end{split}
\eeq

\noindent
This sum is dominated by the second term that describes transitions between the states with zero orbital momentum. We substitute this sum  in  \eq{generalgamma} and obtain $\Gamma(P_c\to\Lambda_c+\bar D^*)=13.3$ MeV.


\begin{figure}[h!]
\begin{center}
\includegraphics[width=4 cm]{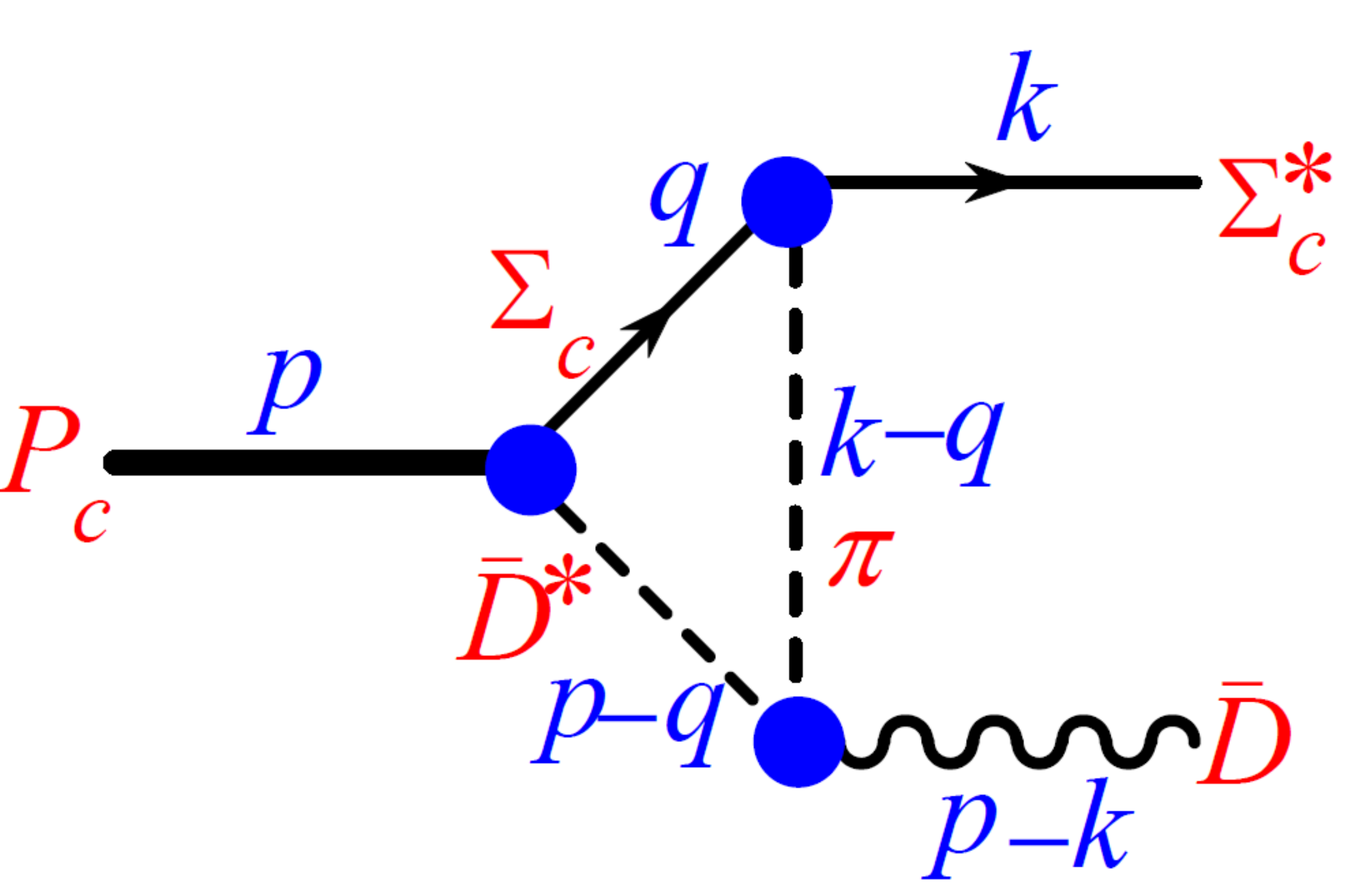}
\end{center}
\caption{Decay of molecular pentaquark $P_c(4450)$ into open charm states $\bar{D}+\Sigma^*_c$}
\label{fig8}
\end{figure}

The $P_c\to \Sigma^*_c+\bar D$ decay  goes via the one-pion exchange diagram  in Fig.~\ref{fig8}. The $\pi\Sigma_c^*\Sigma_c$   interaction Lagrangian and coupling constant are in Table~\ref{pionlgr}.
After calculations we obtain the sum of matrix elements squared

\beq
\begin{split}
{\int\mathllap{\sum}}_f \:
|{\mathcal M}_{i\to f}|^2& = 2\left| M_c\left(2,\frac{3}{2}\biggl|2\right)-M_t\left(0,\frac{3}{2}\biggl|2\right)-M_t\left(2,\frac{1}{2}\biggl|2\right)\right|^2\\
&
 +
2\left| M_c\left(0,\frac{3}{2}\biggl|0\right)+M_t\left(2,\frac{1}{2}\biggl|0\right)-M_t\left(2,\frac{3}{2}\biggl|2\right)\right|^2,
\end{split}
\eeq

\noindent
substitute it in \eq{generalgamma} and calculate the width $\Gamma(P_c\to\Sigma_c^*+\bar D)=0.2$ MeV.

\subsection{Decays into States with Hidden Charm }

\begin{figure}[h!]
\begin{center}
\includegraphics[width=12 cm]{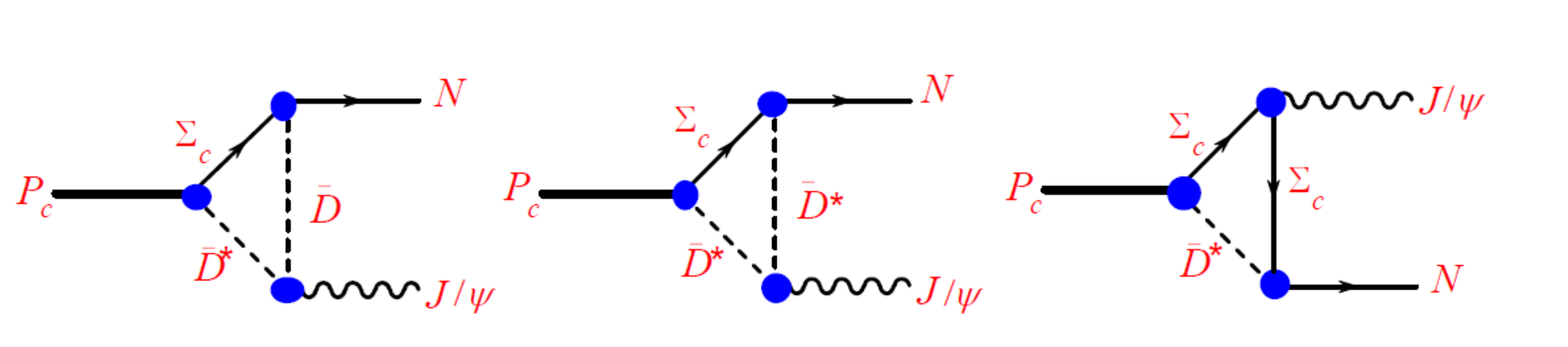}
\end{center}
\caption{Decays of the molecular pentaquark $P_c(4450)$ into hidden charm states $J/\psi+N$}
\label{moldech}
\end{figure}

The $P_c(4450)\to J/\psi+N$ decay  is the only one kinematically allowed two-particle decay of the pentaquark into states with hidden charm. This decay goes via diagrams with exchange by a charmed meson or baryon in $t$-channel, e.g., $D$, $D^*$, $\Sigma_c$, etc. We will account only for the contribution of the diagram in Fig.~\ref{moldech} with the exchange by the lightest charmed particle, the pseudoscalar $D$, that we expect to provide a reasonable estimate of the total decay width. The product of internal parities of $J/\psi$ and $N$ is negative, so decay $P_c(4450)\to J/\psi+N$ goes with the lowest orbital momenta $L=0,2$. The decay momentum $k=820$ MeV in this decay is comparable with the nucleon mass and one cannot use the nonrelativistic approximation for the final nucleon.

\begin{figure}[h!]
\begin{center}
\includegraphics[width=2.5 cm]{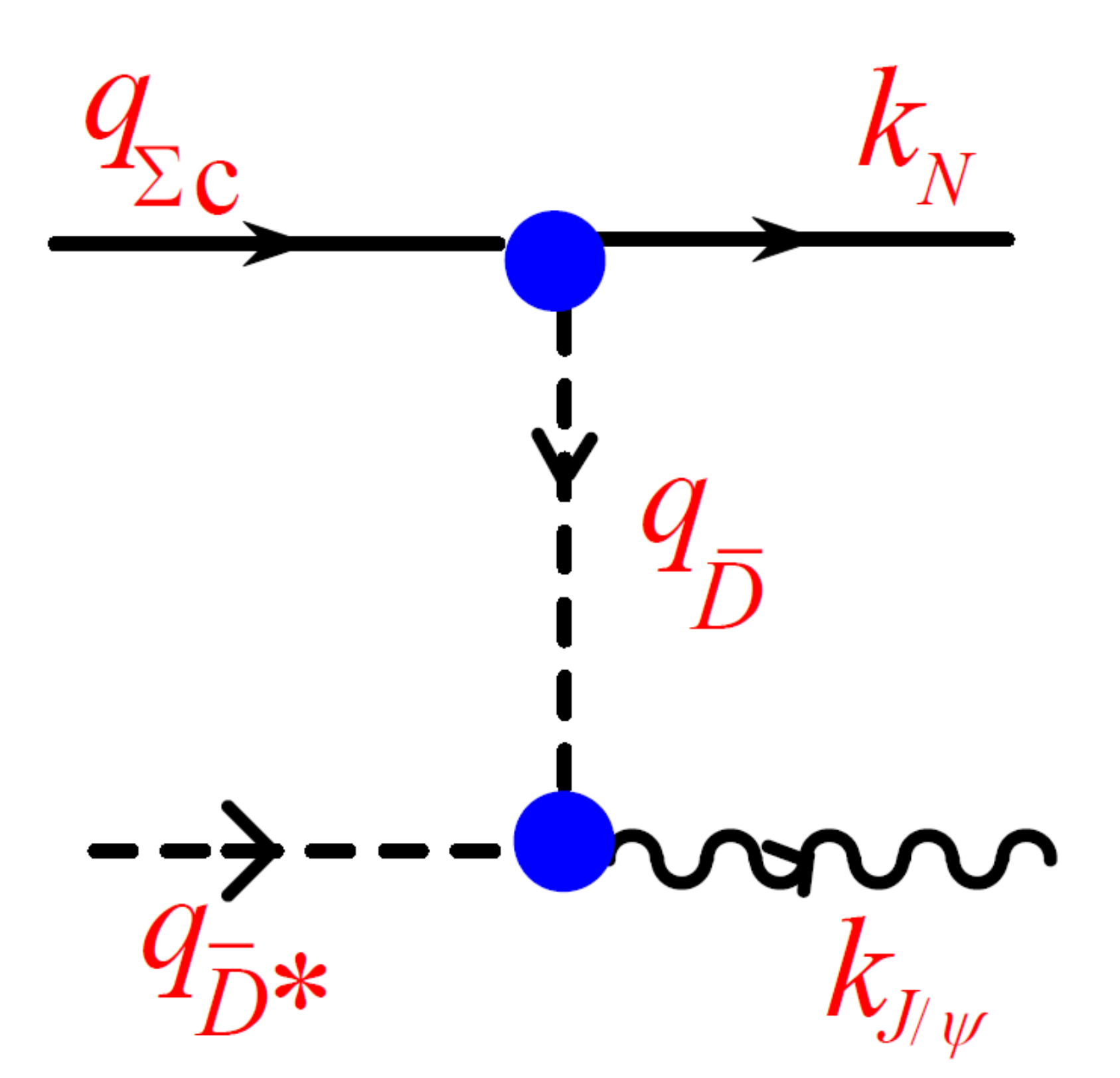}
\end{center}
\caption{Amplitude $\Sigma_c+\bar D^*\to N+J/\psi$}
\label{amplmolhich}
\end{figure}

As with the pion exchanges above, we start with calculation of the relativistic scattering amplitude in Fig.~\ref{amplmolhich}

\beq \label{dexcnhgopcham}
{\mathcal A}(\bm q,\bm k)
=g_{\scriptscriptstyle\Sigma_cDN}g_{\scriptscriptstyle J/\psi D D^*} \epsilon^{*\nu}\bar N(k)\gamma^5\tau^a
\frac{1}{M_{\scriptscriptstyle D}^2-q_{\scriptscriptstyle D}^2}\epsilon^{\mu\nu\alpha\beta}k^{J/\psi}_\mu(q_{\scriptscriptstyle D}-q_{\scriptscriptstyle\bar D^*})_\beta
\Sigma_c^a \bar D^*_\alpha,
\eeq

\noindent
where $\bar D^*_\alpha(\bm q)$ is a four-vector isospinor, $\Sigma^a$ is a spinor isovector,  $N(\bm k)$ is a spinor isospinor, and  $\epsilon^\nu$ is the polarization vector of the final $J/\psi$. The coupling constants and interaction Lagrangians can be found in Table~\ref{fermboslgr} and Table~\ref{bosbosgr}, and are discussed in Appendices~\ref{nuclintlacn} and \ref{charminlagco}.

Next we would like to make a nonrelativistic expansion in the initial momentum $\bm q$.   The denominator of the propagator in \eq {dexcnhgopcham} reduces to $M_*^2(D)+(\bm k-\bm q)^2$ and the range of the effective potential is determined by $M_*(D)=\left[M_{\scriptscriptstyle D}^2-(M_{\scriptscriptstyle \Sigma_c}-E_{\scriptscriptstyle N})^2\right]^\frac{1}{2}=1421$ MeV ($E_{\scriptscriptstyle N}=\sqrt{M_{\scriptscriptstyle N}^2+\bm k^2}$). This effective potential acts at shorter distances  than in the case of the molecular pentaquark  decays into states with open charm. The zero component of the transferred momentum $M_{\scriptscriptstyle \Sigma_c}-\sqrt{M_{\scriptscriptstyle N}^2+k^2}=1208$ MeV is also large. Hence, we cannot neglect the decay momentum and zero component of the transferred momentum in the nonrelativistic limit. As a result the coordinate-dependent term $W_{ik}(\bm r)$ in the transition operator

\beq
\bar{N}^\dagger\sigma_i\Sigma_c^a \tau^a \bar{D}^*_l \epsilon^*_m \varepsilon_{klm} W_{ik}
\eeq

\noindent
is more complicated than the similar term  $W_{ik}(\bm r)$ from \eq{tensorpotcoor} in a fully nonrelativistic case in \eq{potenasopdeld}. In the case at hand

\beq \label{newwikr}
W_{ik}(\bm r) = \delta_{ik}V_c(r) + (3n_in_k-\delta_{ik})V_t(r) +[i( a_1 k_i \partial_k+a_2 k_k \partial_i)
+b k_ik_k] \frac{3V_c(r)}{M_*^2(D)}.
\eeq

\noindent
The derivatives originate from the linear in the relative momentum $\bm q$  terms $q_ik_k$ in the numerator of the momentum space expressions.  Due to these derivatives a new potential

\beq
V_d(r)=\frac{\partial}{\partial r}\left[\frac{3V_c(r)}{M_*(D)}\right]
\eeq

\noindent
arises in $W_{ik}(\bm r)$ in \eq{newwikr} besides the potentials $V_c$ and $V_t$ from \eq{scaltenpot} ($M_*(D)$ plays the role of the mass parameter in all three potentials). We also keep the last bilinear in the final momentum $k_ik_k$ term in the square brackets in \eq{newwikr}  that cannot be legitimately omitted when the final momentum is large. All these new terms are missing in the nonrelativistic decays with exchange by an almost massless pseudogoldstone pion, because its interaction vertex is always proportional to its momentum. But nothing bans such interaction terms for a heavy $\bar D$.

The coefficients in \eq{newwikr} are functions of masses and the final momentum

\beq
a_1=1-\frac{2M_{\scriptscriptstyle \Sigma_c}}{M_{\scriptscriptstyle N}+E_{\scriptscriptstyle N}},\qquad
a_2=\frac{M_{\scriptscriptstyle \Sigma_c}-E_{\scriptscriptstyle N}}{E_{\scriptscriptstyle J/\psi}}, \qquad  b=-a_1 a_2,
\eeq

\noindent
where $E_{\scriptscriptstyle J/\psi}=\sqrt{M_{\scriptscriptstyle J/\psi}+k^2}$ is the energy of the produced $J/\psi$. Notice that these coefficients would be zero if masses of the constituent $\Sigma_c$ and the produced nucleon were close.

Further calculations go almost as in the case of the nonrelativistic decays above. A new element is connected with the scalar products like $\bm k\cdot \bm n$ ($\bm n=\bm r/r$) that arise after differentiation in \eq{newwikr}.  We write them in terms of spherical harmonics $\bm k\cdot \bm n=-i\sqrt{4\pi/3}\sum_m k^{(-m)}Y_{1m}$, where $k^{(-m)}$ are spherical components of $\bm k$.  After application of the transition operator the final wave function contains products of different spherical harmonics that depend on $\bm r/r$ and we use the Clebsch-Gordan coefficients to obtain terms linear in spherical harmonics, integrate over angles with the outgoing plane wave and obtain typical terms $j_L(kr)Y_{LM}(\bm k/k)$. Unlike the decays considered above, now such terms are multiplied  by linear in the spherical components of $\bm k$ factors. We calculate the radial integrals, project each of the products of spherical harmonics of $\bm k/k$ on a single spherical harmonic $Y_{L'M}(\bm k/k)$, square the obtained sums and integrate over directions of $\bm k$. Notice that this calculation leads to the decay products with a final orbital momentum $L'\neq L$ in $M(l,S|L)$ ($L$ is the label of the spherical Bessel function in the respective radial integral). The expression for the sum of matrix elements  squared turns out to be rather cumbersome. The dominant contribution to this sum is supplied by the transitions from the component of the initial bound state wave function  with $l=0$, $S=3/2$ that has the form

\beq
{\int\mathllap{\sum}}_f \:
|{\mathcal M}_{i\to f}|^2 =
3\left(1+\frac{2bk^2}{M_*^2(D)}+\frac{6b^2k^4}{M_*^4(D)}\right)\left|M_c
\left(0,\frac{3}{2}\biggl|0\right)\right|^2
+15\left|M_t\left(0,\frac{3}{2}\biggl|2\right)\right|^2
\eeq
\[
+\frac{30bk^2}{M_*^2(D)} M_t\left(0,\frac{3}{2}\biggl|2\right)M_c\left(0,\frac{3}{2}\biggl|0\right)+ \frac{2\left(a_1+a_2\right)^2k^2}{M^2_*(D)}
\left|M_d\left(0,\frac{3}{2}\biggl|1\right)\right|^2,
\]

\noindent
where we introduced matrix element of a new type

\beq \label{intormomone}
M_d(l,S|L)= \int_0^\infty dr r^2 R_{lS}(r)V_d(r)j_L(kr),
\eeq

\noindent
that arises  only for the odd values of $L$. The potential $V_d(r)$ in this integral is regularized in the same way as the potentials $V_c(r)$ and $V_t(r)$ in \eq{scaltenpot}.

The final nucleon is relativistic in this decay and the general formula for the width in \eq{generalgamma} changes

\beq \label{widthrelpop}
\Gamma=g^2_{\scriptscriptstyle {D}\Sigma_c N}g^2_{\scriptscriptstyle J/\psi{D}{D}^*}\frac{4kE_{\scriptscriptstyle N}
E_{\scriptscriptstyle J/\psi }}{M_{\scriptscriptstyle  P_c}}\frac{E_{\scriptscriptstyle J/\psi}^2}{(2M_{\scriptscriptstyle D^*})(2M_{\scriptscriptstyle \Sigma_c})(2E_{\scriptscriptstyle J/\psi})(2E_{\scriptscriptstyle N})} \frac{E_{\scriptscriptstyle N}+M_{\scriptscriptstyle N}}{2M_{\scriptscriptstyle \Sigma_c}}\: {\int\mathllap{\sum}}_f \:|{\mathcal M}_{i\to f}|^2.
\eeq

\noindent
After numerical calculations we obtain decay width of the molecular pentaquark into states with hidden charm $\Gamma(P_c(4450)\to N+J/\psi)=0.03$ MeV. Account for relativity of the final nucleon significantly affects  this result, the width decreases by 61\% without the relativistic corrections. The suppression of the decay into hidden charm states is somewhat stronger that the one we could expect from the estimates of the matrix elements discussed in the next section. This additional suppression is due to the  small magnitude of the coupling constant $g_{\scriptscriptstyle\Sigma_c ND}$, see  Table~\ref{fermboslgr} and discussion in Appendix~\ref{nuclintlacn}. Let us emphasize that a rather strong suppression due to smallness of the matrix elements would survive even a significant increase of the coupling constant.

\begin{table}[h!]
\caption{\label{pentmoldecop} Pentaquark  $P_c(4450)$ decay widths in the molecular picture}
\begin{ruledtabular}
\begin{tabular}
{lllll}
Decay mode
& $L$\footnote{Lowest allowed orbital momentum.}
&$k$\footnote{Final momentum.}  (MeV)
&
$m_*$\footnote{Effective exchanged mass.}  (MeV)
& $\Gamma$\footnote{Decay width.} (MeV)
\\
$P_c\to \Lambda_c\bar{D}$& 2 & 798 & 136 & 6.8
\\
$P_c\to \Sigma_c\bar{D}$& 2 & 529  & 128 & 1.4
\\
$P_c\to \Lambda_c\bar{D}^*$& 0,2  & 579  & 101  & 13.3
\\
$P_c\to \Sigma^*_c\bar{D}$& 0,2  & 360  & 107 &  0.2
\\
\hline
$P_c\to J/\psi N$ & 0  & 820  &  1421  &   0.03
\\
\hline
Total width &   &   &   &   21.7
\end{tabular}
\end{ruledtabular}
\end{table}

\subsection{Comparison of Molecular Pentaquark Decays  into States with Hidden and Open Charm\label{cpmpmolddecm}}

The results collected in Table~\ref{pentmoldecop} demonstrate that  the decay into states with hidden charm is suppressed in comparison with the decays into states with open charm in the molecular picture. As already mentioned in the Introduction this happens because an exchange by a heavy charmed particle is required in decays to the hidden charm states. Let us recap the arguments given in the Introduction.  We argued that in order to decay into hidden charm state the constituents in the molecular picture should come to a small distance $\sim1/m_c$. This is a tiny scale in comparison with the scale of the wave function $\sim1/\kappa\gg1/m_c$ and therefore this width is proportional to $\int d^3r|\psi(\bm r)|^2\sim|\psi(0)|^2/m_c^3\sim (\kappa/m_c)^3$. For molecular pentaquark $\kappa=\sqrt{2\mu\epsilon}\approx 182$ MeV and $(\kappa/m_c)^3\sim 3\times 10^{-3}$. As we will show below this estimate is too naive and the characteristic distance in molecular decays into states with hidden charm is determined not by $m_c$ but  by the mass of a heavy exchanged particle, with the effective mass $M_*$ that grows only as $\sqrt{m_c}$ with $m_c$.

Let us try to improve the naive estimate of molecular decays into states with hidden charm. Recall that the decay amplitudes are sums of the overlap integrals similar to the ones in \eq{matrelll} and \eq{intormomone}

\beq \label{estimacontr}
M(l,S|L)_{c,d,t}= \int_0^\infty dr r^2 R_{lS}(r)V_{c,d,t}(r)j_L(kr).
\eeq

\noindent
where the potentials are defined in \eq{scaltenpot} and \eq{intormomone}. We collected results of the numerical calculations of matrix elements  $M(l,S|L)$ for a typical decay without charm exchange in Table~\ref{overlap-open} and with charm exchange in Table~\ref{overlap-hidden}, respectively.

\begin{table}[h!]
\caption{\label{overlap-hidden}  Molecular pentaquark decay $P_c\to J/\psi+N$: matrix elements}
\begin{ruledtabular}
\setlength{\tabcolsep}{0.1pt}
{\scriptsize
\begin{tabular}{cccccccccc}
&$M\left(0,\frac{3}{2}\biggl|0\right)$&$M\left(2,\frac{1}{2}\biggl|0\right)$&$M\left(2,\frac{3}{2}\biggl|0\right)$
&$M\left(0,\frac{3}{2}\biggl|1\right)$&$M\left(2,\frac{1}{2}\biggl|1\right)$&$M\left(2,\frac{3}{2}\biggl|1\right)$&
$M\left(0,\frac{3}{2}\biggl|2\right)$&$M\left(2,\frac{1}{2}\biggl|2\right)$&$M\left(2,\frac{3}{2}\biggl|2\right)$
\\
$V_c$&0.0232835$$&$ 1.48\times 10^{-3}$&$ -3.72\times 10^{-3}$
&&&
&$ -4.33\times 10^{-3}$&$ 2.47\times 10^{-4}$&$ -6.44\times 10^{-4}$
\\
$V_t$ & &$-7.10\times 10^{-3}$ &
$-1.74\times 10^{-2}$  &
&&
& $-1.37\times 10^{-2}$&
$6.37\times 10^{-4}$ & $ -1.64\times 10^{-3}$
\\
$V_d$ &&&
&
$7.76\times 10^{-2}$ & $-3.11\times 10^{-3}$ & $7.96\times 10^{-3}$ &&
\end{tabular}
}
\end{ruledtabular}
\end{table}

In decays with charm exchange the effective mass $M_*$ is much larger than the decay momentum $k$ and the scale of the wave function  $\kappa$, $M_*\gg k>\kappa$, see Table~\ref{pentmoldecop}. Then

\beq
M_{c,d,t}(l,S|L)\sim \int_0^{\frac{1}{M_*}} dr r^2 (\kappa r)^l(kr)^LV_{c,d,t}(r)
\sim \left(\frac{\kappa}{M_*}\right)^l\left(\frac{k}{M_*}\right)^L\frac{V_{c,d,t}\left(\frac{1}{M_*}\right)}{M_*^3}.
\eeq

\noindent
The sum $l+L\geq2$ in the integrals with the tensor potential  and the overlap matrix element is at most   $M_{t}\sim (k/M_*)^2$ at $l=0$ and $L=2$. In the integral with the potential $V_d(r)$ $L$ is always odd, and this integral is at most $M_{d}\sim k/M_*$ at $l=0$ and $L=1$. It enters the decay amplitude with an additional factor $k/M_*$ and as a result contributes to the  decay amplitude at most $(k/M_*)^2$, exactly like the tensor potential. Finally, naively the contribution of the central potential $V_c$ to the integral in \eq{estimacontr} at $l=L=0$ seems to be independent of $M_*$ when $M_*$ increases. This contradicts the well grounded physical expectations that exchange by a very massive particle should supply negligible contribute to the decay width. It is not hard to figure out what happened. Calculating the Fourier transform in \eq{tensorpotcoor} we have thrown away the $\delta$-function term as unphysical in the case of exchange by a light pion. However, the calculation above shows that for a heavy exchange this $\delta$-function is necessary to restore the proper dependence of the $l=L=0$ decay matrix element on mass of the exchanged particle.  It is easy to see that restoration of   $\delta$-function reduces to substitution $M_{c}(0,S|0)\to M_{c}(0,S|0)-R_{0S}(0)/(12\pi)$. We made this subtraction in calculations of all molecular and hadrocharmonium decays with charm exchange. The subtracted matrix elements are at most $(k/M_*)^2$ and we conclude that effectively all matrix elements in  \eq{estimacontr} decrease with $M_*$ as $(k/M_*)^2$ or faster.

\begin{table}[h!]
\caption{\label{overlap-open}  Molecular pentaquark decay $P_c\to \Lambda_c+\bar D^*$: matrix elements}
\begin{ruledtabular}
\begin{tabular}{ccccccc}
&$M\left(0,\frac{3}{2}\biggl|0\right)$&$M\left(2,\frac{1}{2}\biggl|0\right)$&$M\left(2,\frac{3}{2}\biggl|0\right)$&
$M\left(0,\frac{3}{2}\biggl|2\right)$&$M\left(2,\frac{1}{2}\biggl|2\right)$&$M\left(2,\frac{3}{2}\biggl|2\right)$\\
$V_c$&$-1.95\times 10^{-3}$&$ 1.09\times 10^{-4}$&$ -2.86\times 10^{-4}$
&$ -8.90\times 10^{-4}$&$ 6.97\times 10^{-5}$&$ -1.86\times 10^{-4}$
\\
$V_t$ & &$1.36\times 10^{-2}$ &$-3.43\times 10^{-2}$ & $-2.96\times 10^{-2}$ & $2.00\times 10^{-3}$ & $ -5.30\times 10^{-3}$
\\
\end{tabular}
\end{ruledtabular}
\end{table}

Molecular decays into open charm states go via exchange by the light pion, only the potentials $V_{c,t}$ give contribution to these decays, and $m_*\sim m_\pi$. Numerically, in this case $m_*\sim\kappa\ll k$ Then integration in \eq{estimacontr} goes up to $r\sim 1/k\ll1/\kappa\sim 1/m_*$ and

\beq
M(l,S|L)_{c,t}\sim \int_0^{\frac{1}{k}} dr r^2 (\kappa r)^lV_{c,t}(r)j_L(kr).
\eeq

\noindent
In this region the matrix element of the scalar potential $M(l,S|L)_{c}\sim (\kappa/k)^l(m_*/k)^2$ is suppressed in comparison with the matrix element of the tensor potential $M(l,S|L)_{t}\sim (\kappa/k)^l\sim (m_*/k)^l$ by the factor $(k/m_*)^2\sim 15-30$.

Now we can estimate ratio $R$ of  matrix elements for decay into states with hidden and open charm

\beq
R\sim\left(\frac{k_{hid}}{M_*}\right)^L\biggl/\left(\frac{m_*}{k_{open}}\right)^l,
\eeq

\noindent
where  $k_{open}$ and $k_{hid}$ are decay momenta in the hidden and open charm decays, respectively, and $M_*=M_*(D)$. We compare matrix elements for hidden charm decays with the tensor matrix elements in open charm decays since scalar matrix elements in open charm decays are suppressed. Numerically for decays in Table~\ref{overlap-hidden} and Table~\ref{overlap-open} $R\sim 0.4^l\times 0.5^L\sim 0.1-0.2$. Respectively, we expect that the hidden charm decays of the molecular pentaquark should be suppressed by a factor $0.01-0.04$, what is compatible with the results in Table~\ref{pentmoldecop}. This suppression is weaker than the naive suppression factor $(\kappa/m_c)^3\sim 10^{-6}$ discussed above.

\section{Hadrocharmonium Decays}

\subsection{Decays into States with Hidden Charm }

In the hadrocharmonium picture the LHCb pentaquark $P_c(4450)$ is interpreted as a bound state of $\psi'$ and the nucleon \cite{epp2016,epp2018} (see also \cite{abfs2018}).  It is described by a nonrelativistic wave function that is a product of the $S$-wave coordinate wave function and the spin $3/2$ and isospin $1/2$ factor. The partial decay width  of the hadrocharmonium pentaquark $\Gamma(P_c(4450)\to J/\psi+N)\approx 11$ MeV was calculated in \cite{epp2016,epp2018}. As mentioned above this is the only one kinematically allowed two-particle pentaquark decay channel into states without open charm.

\subsection{Decays into States with Open Charm }

Hadrocharmonium decays into states with open charm go via exchange by  heavy  hadrons. As in the molecular decays we will take into account only exchanges by the lightest particle with open charm, namely by $D$-meson. We expect that the respective partial widths are reasonably well approximated by this exchange. The inverse size $\kappa=\sqrt{2\mu\epsilon}=506$ MeV of the hadrocharmonium pentaquark wave function is determined by  its binding energy $\epsilon=178$ MeV and reduced mass $\mu=720$ MeV. Recall that in the case of the molecular pentaquark we obtained $\kappa=182$ MeV. Hence, the hadrocharmonium wave function is less extended and is larger at the origin than the molecular one. This favors decays with exchange of charm and one can expect that the hadrocharmonium decays into states with open charm have larger partial widths than the molecular pentaquark decay into $J/\psi N$. It is harder to anticipate relative magnitude of partial decay widths into states with open charm in the hadrocharmonium and molecular pictures. On the one hand larger at the origin and less extended hadrocharmonium wave function  could probably enhance decay rates into the four channels with open charm. On the other hand the effective masses of the exchanged particles in these decays are much higher than in the case of the molecular pentaquark (compare Tables~\ref{pentmoldecop} and \ref{penthadrdec}), what works in the opposite direction. Only calculations will show which effect is more pronounced.

\subsubsection{$P_c\to \Lambda_c+\bar{D}$}

\begin{figure}[h!]
\begin{center}
\includegraphics[height=2.cm]{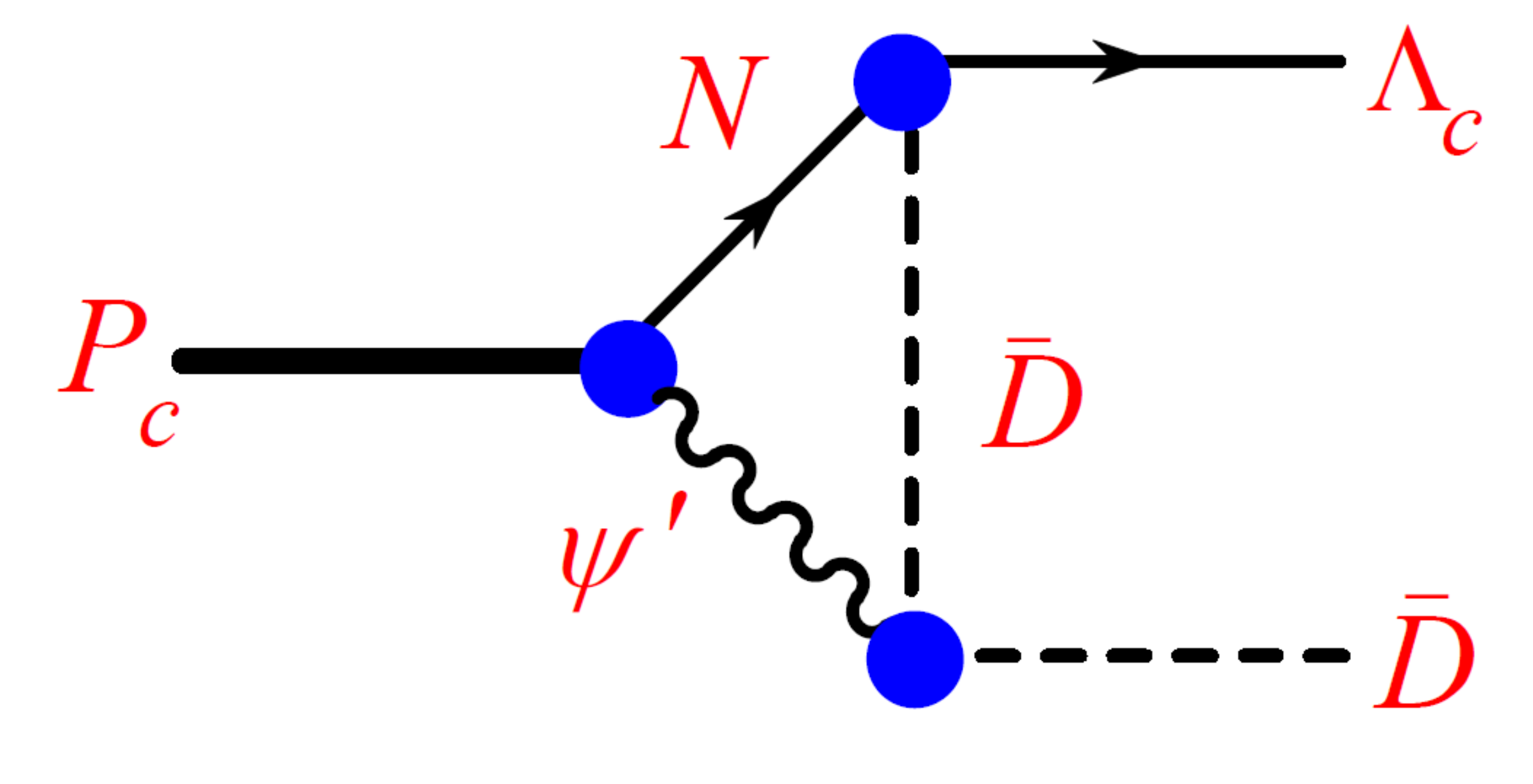}
\end{center}
\caption{Decay of  hadrocharmonium pentaquark $P_c(4450)$ into states with open charm $\Lambda_c+\bar D$
\label{opnechhadr}}
\end{figure}

Consider first the hadrocharmonium decay $P_c\to \Lambda_c+\bar{D}$.  Kinematics of this decay was already discussed above. This decay can go via exchange by the $D$-meson and heavier particles with open charm. As already explained we calculate the partial decay width due to the diagram with the pseudoscalar $D$ exchange in Fig.~\ref{opnechhadr} and expect that this exchange provides a reasonable estimate of the total partial decay width into $\Lambda_c$ and $\bar{D}$.

\begin{figure}[h!]
\begin{center}
\includegraphics[height=3.cm]{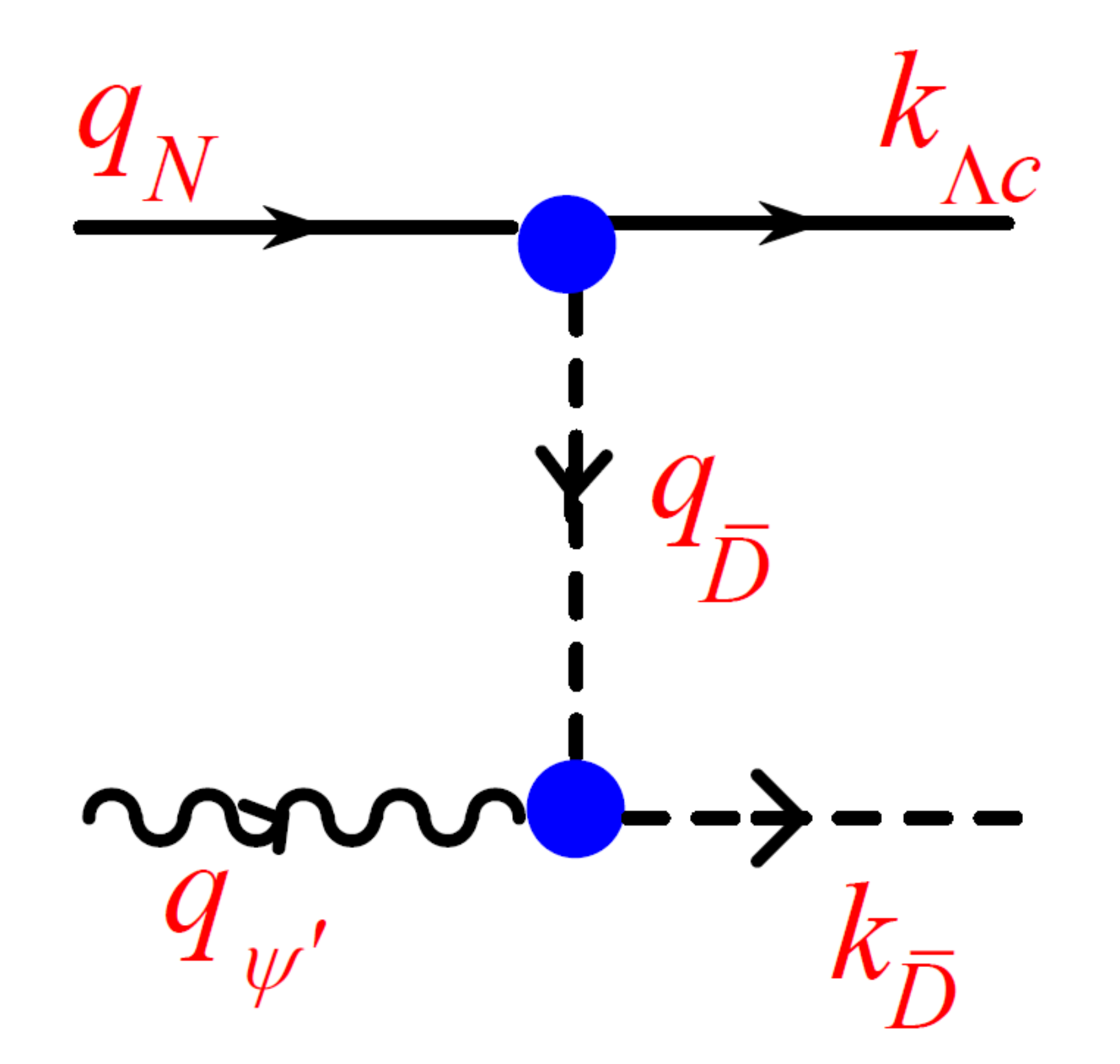}
\end{center}
\caption{Amplitude $N+\psi'\to \Lambda_c+\bar D$
\label{scatnpsplambd}}
\end{figure}

As usual we first calculate  the relativistic scattering amplitude $N+\psi'\to \Lambda_c+\bar D$ in Fig.~\ref{scatnpsplambd} (momenta are labeled as in the figure)

\beq \label{relamphadrtlamd}
\mathcal{A}(\bm q,\bm k)
=
g_{\scriptscriptstyle\Lambda_cDN}g_{\scriptscriptstyle\psi'DD} \bar \Lambda_c(\bm k_{\scriptscriptstyle\Lambda_c})\gamma^5\bar D^\dagger
\frac{1}{M_{\scriptscriptstyle D}^2-q_{\scriptscriptstyle eD}^2}
(q_{\scriptscriptstyle\bar D}+k_{\scriptscriptstyle\bar D^*})^\alpha\Phi_\alpha N,
\eeq

\noindent
where $\Phi_\alpha(\bm q)$ is a four-vector that describes initial $\psi'$, $N$ is a spinor isospinor, $D$ is an isospinor, and  $\Lambda_c(\bm k_{\scriptscriptstyle\Lambda})$ is a spinor.The isospin indices are contracted along the virtual $\bar D$ line. The coupling constants and interaction Lagrangians are collected in Table~\ref{fermboslgr} and Table~\ref{bosbosgr} and discussed in Appendices~\ref{nuclintlacn} and \ref{charminlagco}.

In the nonrelativistic expansion in the initial momentum $\bm q$ the denominator of the propagator in \eq{relamphadrtlamd} reduces to $M_*^2(D)+(\bm k-\bm q)^2$ and the range of the effective potential is determined by
$M_*(D)=\left[M_{\scriptscriptstyle D}^2-(E_{\scriptscriptstyle \Lambda_c}-M_{\scriptscriptstyle N})^2\right]^\frac{1}{2}\approx1133$ MeV ($E_{\scriptscriptstyle \Lambda_c}=(M_{\scriptscriptstyle \Lambda_c}^2+\bm k^2)^\frac{1}{2}$). The relativistic amplitude in the nonrelativistic limit reduces to the transition operator

\beq
(\Lambda_c^\dagger \sigma^i N^a)W_{ik}(\bm r)(D^{\dagger a}\psi'_k),
\eeq

\noindent
where $W_{ik}(\bm r)$ has the same form as  in \eq{newwikr} with the natural kinematic substitutions and

\beq
a_1=1-\frac{2M_{\scriptscriptstyle N}}{E_{\scriptscriptstyle \Lambda_c}+M_{\scriptscriptstyle \Lambda_c}},\qquad
a_2=-1,\qquad b=-a_1a_2.
\eeq

\noindent
We preserved the external momentum $k$ in the transition operator.  Next we apply the transition operator to the initial wave function (compare \eq{fintwafrlamd}) and calculate the sum of matrix elements squared of the transition amplitude  (compare \eq{summelsqld})

\beq
\begin{split}
{\int\mathllap{\sum}}_f \:
|{\mathcal M}_{i\to f}|^2& =
 3\left|M_t\left(0,\frac{3}{2}\biggl|2\right)\right|^2 +\frac{(a_1+a_2)^2k^2}{3M_*^2(D)}\left|
M_d\left(0,\frac{3}{2}\biggl|1\right)\right|^2
\\
&+\frac{3b^2k^4}{M_*^4(D)}\left|M_c\left(0,\frac{3}{2}\biggl|0\right)\right|^2
+\frac{6bk^2}{M_*^2(D)}M_c\left(0,\frac{3}{2}\biggl|0\right)M_t\left(0,\frac{3}{2}\biggl|2\right).
\end{split}
\eeq

\noindent
The partial decay  width is (compare \eq{widthrelpop})

\beq \label{hadrdeclamd}
\begin{split}
\Gamma(P_c\to \Lambda_c+\bar{D})& =g^2_{\scriptscriptstyle \Lambda_c D N}g^2_{\scriptscriptstyle \psi'DD}\frac{4kE_{\scriptscriptstyle \Lambda_c}
E_{\scriptscriptstyle D}}{M_{\scriptscriptstyle P_c}}\frac{1}{(2M_{\scriptscriptstyle N})(2M_{\psi'})(2E_{\scriptscriptstyle \Lambda_c})(2E_{\scriptscriptstyle D})}\frac{M_{\scriptscriptstyle \Lambda_c}+E_{\scriptscriptstyle \Lambda_c}}{2M_{\scriptscriptstyle N}}
\\
&\times{\int\mathllap{\sum}}_f \: |{\mathcal M}|^2_{i\to
f}\approx 0.6~\mbox{MeV}
\end{split}
\eeq

\subsubsection{Other Open Charm Decays of Hadrocharmonium Pentaquark }

Calculations of other three decays of the hadrocharmonium pentaquark into the open charm states

\beq
P_c\to \Sigma_c+\bar D,\qquad P_c\to \Lambda_c+\bar D^*,\qquad P_c\to \Sigma^*_c+\bar D,
\eeq

\noindent
are similar to the calculations above. All these decays go via exchange by the lightest particle with an open charm, $D$-meson. Kinematics for all these decays was already considered above and we will not repeat this discussion.

\begin{figure}[h!]
\begin{center}
\includegraphics[width=4 cm]{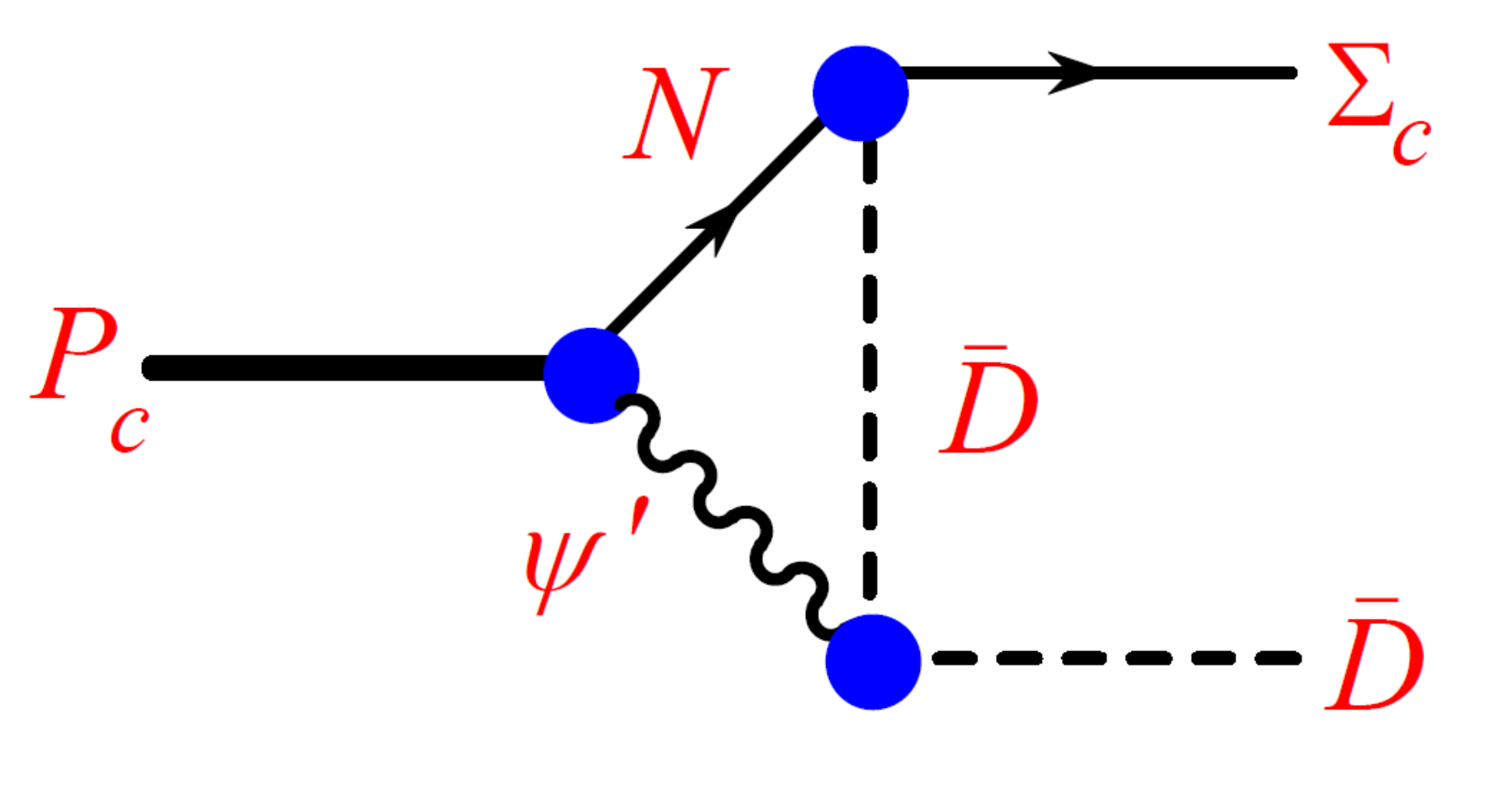}
\end{center}
\caption{Decay of  hadrocharmonium pentaquark $P_c(4450)$ into states with open charm $\Sigma_c+\bar D$}
\label{pcsigmcbd}
\end{figure}

The $P_c\to \Sigma_c+\bar D$ decay is described by the $D$-exchange diagram in Fig.~\ref{pcsigmcbd}, that is similar to the $D$-exchange for $P_c\to \Lambda_c+\bar D$. Effective mass of the exchanged $D$-meson in this decay is $M_*(D)=1005$ MeV. The amplitude for this decay differs from the decay $P_c\to \Lambda_c+\bar{D}$ only by the isospin factor that generates an enhancement factor $3$ in the width. On the other hand the relationship between the coupling constants  $g_{\scriptscriptstyle \Sigma_c N D }=g_{\scriptscriptstyle \Lambda_c N D }/(3\sqrt{3})$ (see \eq{lamcsigmscr} in Appendix \ref{nuclintlacn}) supply a suppression factor for the  $P_c\to \Sigma_c+\bar D$ decay. After replacement  of the coupling constants, masses and multiplication by $3$ we can use \eq{hadrdeclamd} for calculation of the $P_c\to \Sigma_c+\bar D$ partial decay width.  We obtain $\Gamma(P_c\to \Sigma_c+\bar D) =0.036$ MeV, see Table~\ref{penthadrdec}. The suppression  by an order of magnitude $\sim 1/9$ relative to the decay $P_c\to \Lambda_c+\bar D$ comes mainly from the ratio of the coupling constants squared times three from the isotopic factor, difference between the masses of $\Sigma_c$ and $\Lambda_c$ plays an insignificant role.

\begin{figure}[h!]
\begin{center}
\includegraphics[width=4 cm]{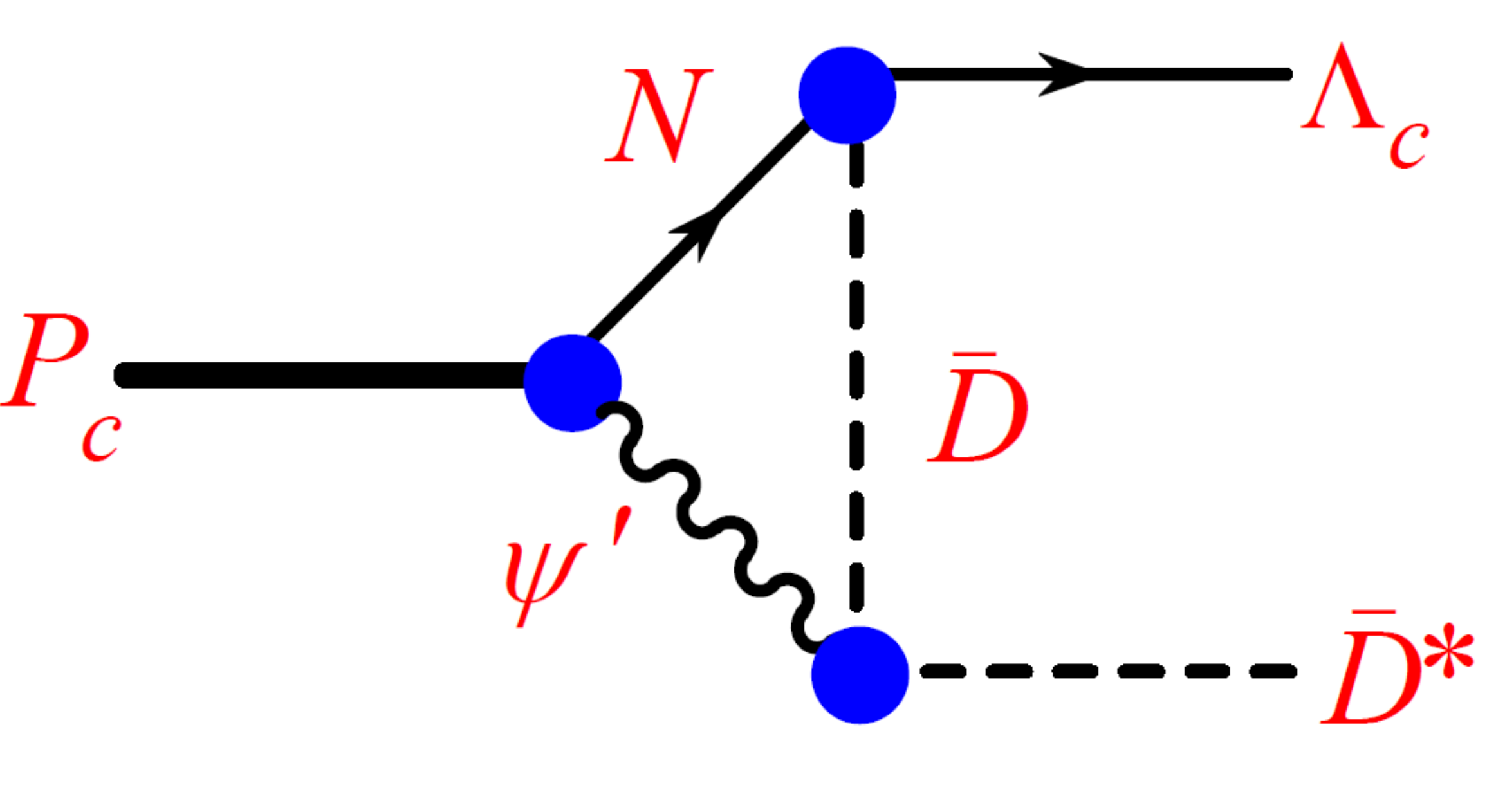}
\end{center}
\caption{Decay of  hadrocharmonium pentaquark $P_c(4450)$ into states with open charm $\Sigma_c+\bar D^*$}
\label{hdecaydstr}
\end{figure}

To calculate the partial decay width $P_c\to \Lambda_c+\bar D^*$ (see Fig.~\ref{hdecaydstr}) we go through the by now standard steps: calculate the relativistic scattering amplitude $N+\psi'\to \Lambda_c+\bar D^*$, make the nonrelativistic approximation for the constituent hadrons, derive an expression for the transition operator and calculate the decay amplitude. The sum of the matrix elements squared for the decay $P_c\to \Lambda_c+\bar D^*$ turns out to be

\beq
\begin{split}
{\int\mathllap{\sum}}_f \:
|{\mathcal M}|^2_{i\to f}&= \left|M_c\left(0,\frac{3}{2}\biggl|0\right)\right|^2+5\left|M_t\left(0,\frac{3}{2}\biggl|2\right)\right|^2 +
\frac{2(a_1+a_2)^2k^2}{3M_*^2(D)}\left| M_d\left(0,\frac{3}{2}\biggl|1\right)\right|^2
\\
&+\frac{2b(1+3b)k^4}{M_*^4(D)}\left|M_c\left(0,\frac{3}{2}\biggl|0\right)\right|^2
+\frac{10bk^2}{M_*^2(D)}M_c\left(0,\frac{3}{2}\biggl|0\right)M_t\left(0,\frac{3}{2}\biggl|0\right),
\end{split}
\eeq

\noindent
where

\beq
a_1=\frac{M_{\scriptscriptstyle N}-E_{\scriptscriptstyle\Lambda_c}}{M_{psi'}+M_{\scriptscriptstyle N}-M_{\scriptscriptstyle\Lambda_c}}, \qquad a_2=
1-\frac{2M_{\scriptscriptstyle N}}{M_{\scriptscriptstyle N}+E_{\scriptscriptstyle\Lambda_c}}, \qquad b=-a_1 a_2.
\eeq

\noindent
The partial width is

\beq
\begin{split}
\Gamma(P_c\to \Lambda_c+\bar D^*)& =
g^2_{\scriptscriptstyle \Lambda_c ND}g^2_{\scriptscriptstyle \psi' DD^*}\frac{4kE_{\scriptscriptstyle \Lambda_c}
E_{\scriptscriptstyle D^*}}{M_{\scriptscriptstyle P_c}}\frac{(M_{\psi'}+M_{\scriptscriptstyle N}-E_{\scriptscriptstyle \Lambda_c})^2}{(2M_{\scriptscriptstyle N})(2M_{\psi'})(2E_{\scriptscriptstyle \Lambda_c})(2E_{\scriptscriptstyle D^*})}
\frac{M_{\scriptscriptstyle \Lambda_c}+E_{\scriptscriptstyle \Lambda_c}}{2M_{\scriptscriptstyle N}}
\\
&\times{\int\mathllap{\sum}}_f \: |{\mathcal M}|^2_{i\to f}
\approx 4.2~\mbox{MeV}.
\end{split}
\eeq


\begin{figure}[h!]
\begin{center}
\includegraphics[width=4 cm]{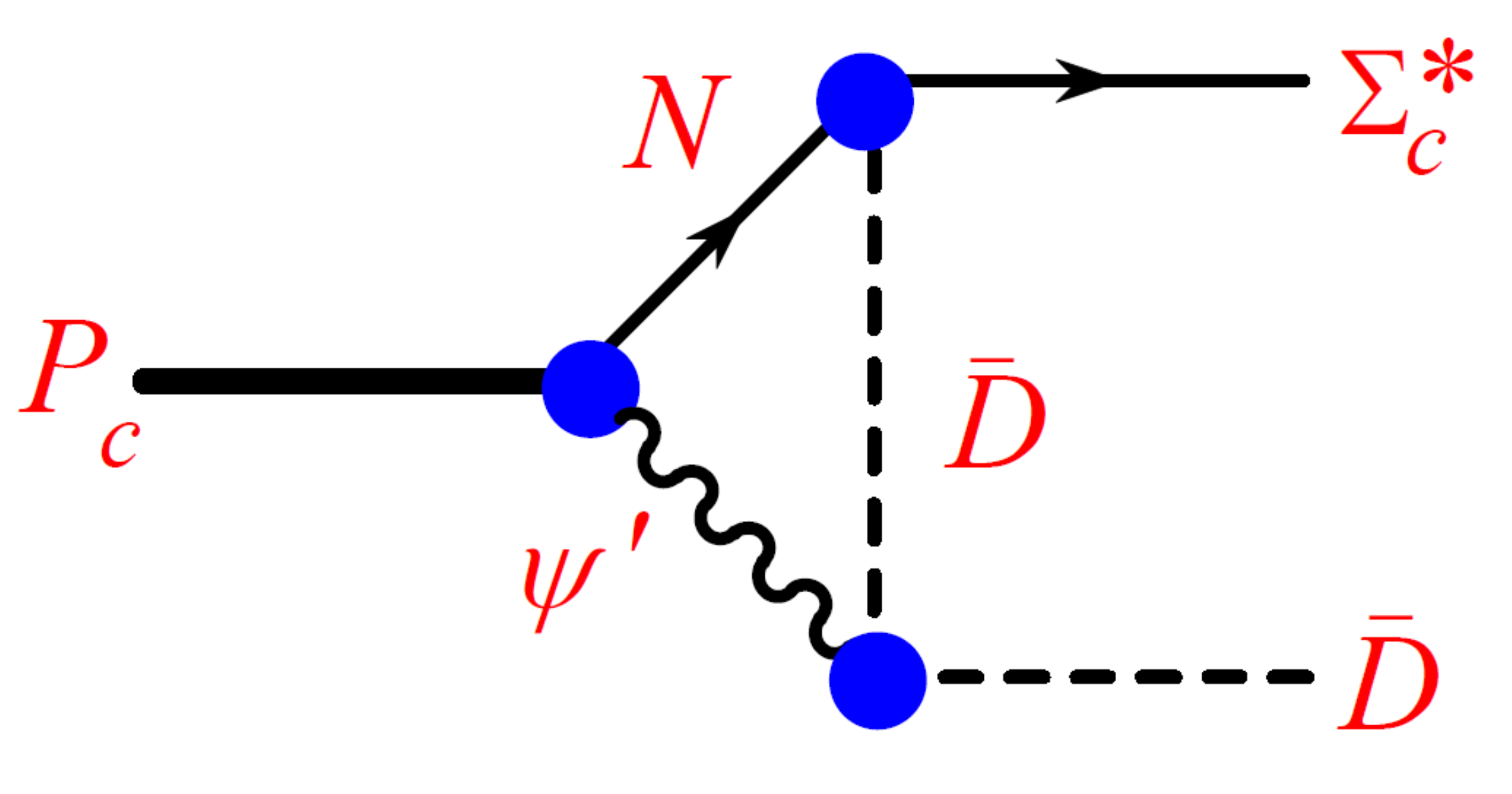}
\end{center}
\caption{Decay of hadrocharmonium pentaquark $P_c(4450)$ into states with open charm $\bar{D}+\Sigma^*_c$}
\label{hadrsigstbd}
\end{figure}

\begin{figure}[h!]
\begin{center}
\includegraphics[height=2.5cm]{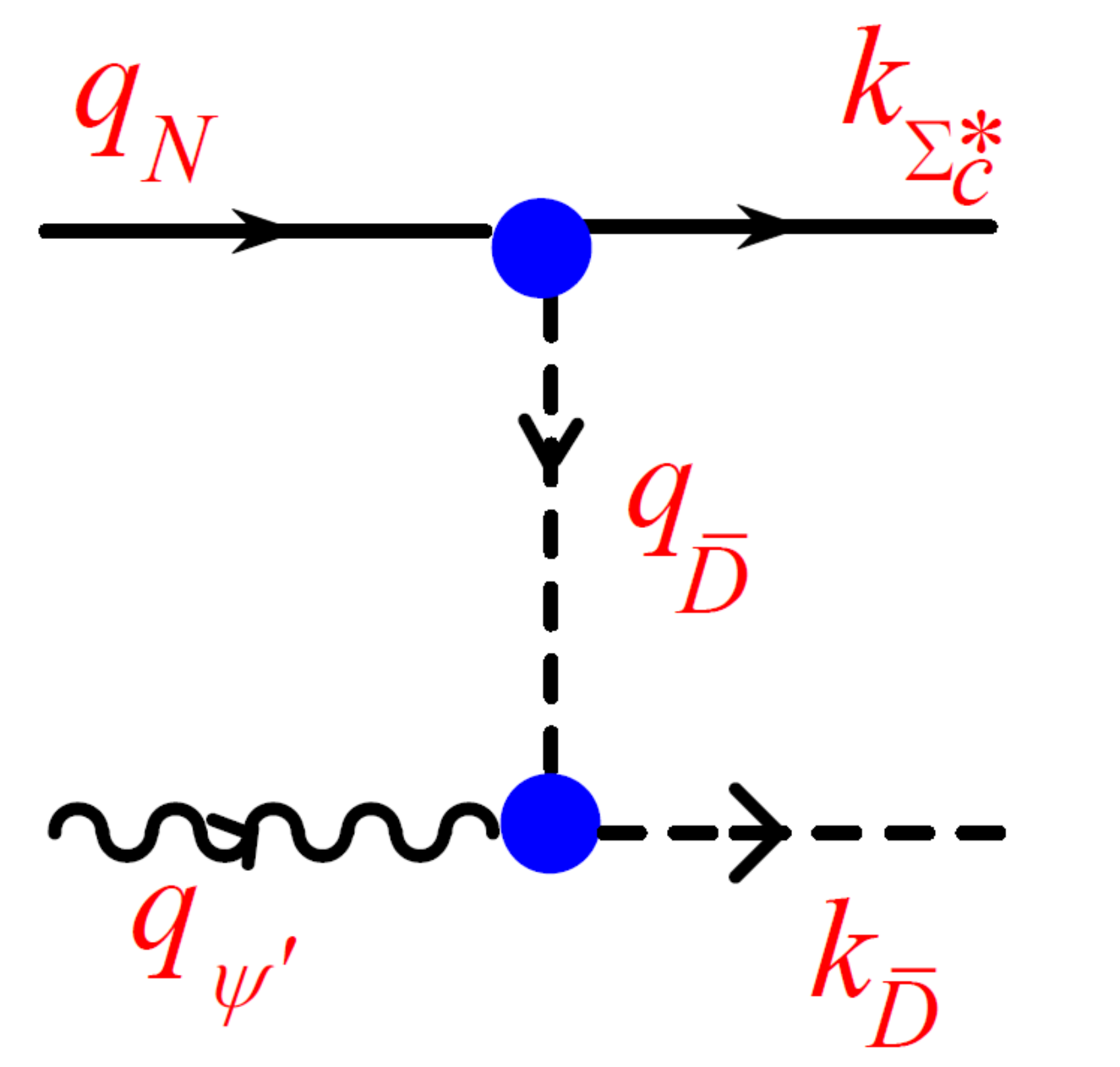}
\end{center}
\caption{Amplitude $N+\psi'\to \Sigma^*_c+\bar D$\label{hadrdecdsst}}
\end{figure}

The $P_c\to \Sigma_c^*+\bar D$ decay  goes via the $D$-exchange diagram  in Fig.~\ref{hadrsigstbd}. The $\Sigma^*N D$  interaction Lagrangian (notice absence of $\gamma^5$!) is in Table~\ref{fermboslgr}. We again go through the standard steps: calculate the relativistic scattering amplitude in Fig~\ref{hadrdecdsst}, use this amplitude with the nonrelativistic initial particles to derive the transition operator, obtain the decay amplitude, sum matrix elements squared and calculate the decay width  $\Gamma(P_c\to \Sigma_c^*+\bar{D})=0.42$ MeV.

\begin{table}[h!]
\caption{\label{penthadrdec} Pentaquark  $P_c(4450)$ decay widths in the hadrocharmonium picture }
\begin{ruledtabular}
\begin{tabular}
{lcccc}
Decay mode
& $L$\footnote{Lowest allowed orbital momentum.}
&$k$\footnote{Final momentum.}  (MeV)
&
$M_*(D)$\footnote{Effective exchanged mass.}  (MeV)
& $\Gamma$\footnote{Decay width.} (MeV)
\\
\hline
$P_c\to J/\psi N$ & 0  & 820  &   & $11$
\\
\hline
$P_c\to \Lambda_c\bar{D}$& 2 & 798 & 1133 &  $0.6$
\\
$P_c\to \Sigma_c\bar{D}$& 2 & 529  & 1005 & 0.04
\\
$P_c\to \Lambda_c\bar{D}^*$& 0,2  & 579  & 1218  & 4.2
\\
$P_c\to \Sigma^*_c\bar{D}$& 0,2  & 360  & 959 & 0.4
\\
\hline
Total width&&&&16.2\
\end{tabular}
\end{ruledtabular}
\end{table}

\section{Discussion of Results}

We calculated the total and partial decay widths of the hadrocharmonium and molecular pentaquarks $P_c(4450)$ constructed in \cite{epp2016,epp2018}.  One could expect that decays into states with open charm dominate in the case of the molecular pentaquark, while the decay to $J/\psi N$ would be the dominant mode for the hadrocharmonium pentaquark, see discussion in the Introduction. The calculations above confirm these expectations both for the molecular and hadrocharmonium pentaquarks, see Tables~\ref{pentmoldecop} and \ref{penthadrdec}. Total decay widths of the molecular and hadrocharmonium pentaquarks are comparable and are about a few dozen MeV in  both scenarios. Taking into account uncertainties of the phenomenological coupling constants and unaccounted for relativistic corrections to the semirelativistic approximation used in the calculations these total widths are comfortably compatible with the  width $\Gamma=39\pm5\pm 19$ MeV measured experimentally \cite{LHCb2015,LHCb2016}.

We expect that the results for the relative magnitudes of partial decays widths in different open channels are more reliable than their absolute values. This happens because in the ratios of the partial widths values of the poorly known interaction constants often cancel and the ratios are more dependent on the matrix elements of the perturbation potentials between the initial and final wave functions. The partial decay width of the molecular pentaquark into the hidden charm states $J/\psi N$ is strongly suppressed, it is about one, two or three three orders of magnitude smaller than the partial widths for decays into different channels with open charm, see Table~\ref{pentmoldecop}\footnote{Recent nonobservation of the pentaquark resonance in the formation reaction $\gamma+p\to J/\psi+p$ \cite{cluex2018} could be interpreted as an indication of the molecular nature of the LHCb pentaquark. However, it is hard to reconcile this result with the initial LHCb discovery of the pentaquark in the invariant mass distribution of $J/\psi N$. Clearly more work is needed and it is too early to come to any definite conclusions.}. The suppression can be understood if we recall that the molecular pentaquark has a relatively large size, its root means square radius is about $1.5$ fm \cite{epp2018}. To decay into states with hidden charm constituents of the molecular  pentaquark need to exchange by a heavy charmed meson. In other words they should come very close to one another what is impeded by the large size of the loosely bound state wave function. The detailed considerations of the matrix elements in Section~\ref{cpmpmolddecm} provide a quantitative justification for these conclusions.

Decay pattern of the hadrocharmonium pentaquark also looks like expected. The hadrocharmonium decays into states with open charm are suppressed in comparison with the hadrocharmonium decays into hidden charm states. Quantitatively this suppression is weaker than the suppression of the hidden charm decays in the case of the molecular pentaquark, compare the results in Tables~\ref{pentmoldecop} and \ref{penthadrdec}. One of the partial widths for hadrocharmonium decay into open charm states ($P_c\to\Lambda_c\bar D$) is only two and a half times smaller than the partial decay width to $J/\psi \bar D^*$. To decay into states with open charm constituents in the hadrocharmonium should come close to one another what happens when they exchange by a heavy charmed meson. The relatively weaker suppression of such hadrocharmonium processes in comparison with the respective molecular case decays is due to a larger binding energy and respectively smaller size (about $0.5$ fm) of the hadrocharmonium bound state.

We see that the decay patterns of the molecular and hadrocharmonium pentaquarks are vastly different. In the molecular scenario  decays into $J/\psi$ are strongly suppressed, while  the opposite happens in the hadrocharmonium case when a less pronounced suppression of decays into states with open charm is predicted. Total decay widths are comparable in both scenarios and are about a few dozen MeV. Comparison of these decay patterns with the experimental data would hopefully help to reveal which of the two theoretical scenarios  for pentaquarks (if either) is chosen by nature.

\acknowledgments

This paper was supported by the NSF grant PHY-1724638.

\appendix

\section{Interaction Lagrangians and interaction constants}

A number of phenomenological interaction Lagrangians was used in calculations in the main body of this paper. Coupling constants in these Lagrangians were discussed in the literature many times, see, e.g., \cite{lsgz2017,sgxz2016,hpkmw2017,tmyhyc1992,pc11992,ysdsmh2016,dorgeb2001,lo2012,gz2015,cd2016,wlcmkzwl2001,clm2013,yocmkkn2008,dhhkm2011,fsnmn1998,fsnmn1999,fodfsnmn2001,mebmcfsn2012} and references therein. There is no universal agreement on the values of some of these constants, while decay widths obtained above critically depend on these values. There are three groups of relevant Lagrangians that describe: 1) pion interaction with charmed hadrons, 2)  $D$-boson interactions with baryons, and 3)  $D$-boson interaction with heavy mesons. The interaction Lagrangians and coupling constants are collected in Tables~\ref{pionlgr}-\ref{bosbosgr}. The interaction constants in these tables are known with vastly different degree of reliability. We tried to use the value of this or that constant obtained with a minimal number of theoretical assumptions. Below we discuss how these values arise  and how accurate they are.

\subsection{Pion interaction constants in Table~\ref{pionlgr}\label{pionintrco}}

\begin{table}[h!]
\caption{\label{pionlgr} Pion interactions}
\begin{ruledtabular}
\begin{tabular}
{lll}
Interacting particles &Interaction Lagrangian & Coupling constant
\\
\hline
$\pi\Sigma_c\Lambda_c$& $-ig_{\scriptscriptstyle\pi\Sigma_c\Lambda_c}\bar \Lambda_c^\dagger\gamma^5\bm\Sigma_c\cdot\bm\pi+H.c$
&$g_{\scriptscriptstyle\pi\Sigma_c\Lambda_c}=19.2$ \footnote{From $\Gamma_{exp}(\Sigma^{++}_c\to\Lambda_c\pi^+)=1.89^{+0.09}_{-0.18}$ and $\Gamma(\Sigma_c^{0}\to \Lambda_c\pi^-)=1.83^{+0.11}_{-0.19}$~MeV, see Appendix A.1 and \cite{lsgz2017}.}
\\
$\pi\Sigma_c\Sigma_c$ & $-ig_{\scriptscriptstyle\pi\Sigma_c\Sigma_c}\epsilon_{abc}\bar{\Psi}^a_\Sigma\gamma_5\Psi^b_\Sigma\pi^c+ H.c.$ &
$g_{\scriptscriptstyle\pi\Sigma_c\Sigma_c}=11.06$ \footnote{See Appendix A.1.}
\\
$\pi\Sigma_c\Sigma_c^*$ & $i\tilde g_{\scriptscriptstyle\pi\Sigma_c\Sigma^*_c}\bar\Sigma_c^{*\mu,a}\epsilon_{abc}\Sigma_c^b\partial_\mu\pi^c+ H.c.$ &  $\tilde g_{\scriptscriptstyle\pi\Sigma_c\Sigma^*_c}=\frac{9.7}{\sqrt{2M_{\scriptscriptstyle \Sigma_c^*}}\sqrt{2M_{\scriptscriptstyle \Sigma_c}}}$\footnote{See Appendix A.1.}
\\
$\pi DD^*$ &$ig_{\scriptscriptstyle\pi DD^*} \left(D^{*\dagger}_\mu\partial^\mu\pi D-D^{\dagger}\partial^\mu\pi D^*_\mu\right)$ &
$g_{\scriptscriptstyle\pi D  D^*}=12.12$\footnote{From $\Gamma_{exp}(D^{*+}(2010)\to D^0\pi^+)=56.5\pm0.1$~keV and $\Gamma_{exp}(D^{*+}(2010)\to D^+\pi^0)=25.6\pm0.6$~keV, See Appendix A.1 and \cite{sgxz2016}.}
\\
$\pi D^*D^*$ & $g_{\scriptscriptstyle\pi D^*D^*}\epsilon^{\mu\nu\alpha\beta}D^{*\dagger}_\mu\partial_\nu\pi\partial_\alpha D^{*}_\beta$&
$g_{\scriptscriptstyle\pi D^*D^*}=6.25$ GeV$^{-1}$

\end{tabular}
\end{ruledtabular}
\end{table}

Pion interactions with heavy baryons and mesons are usually described in the framework of the heavy quark effective theory combined with the spontaneously broken $SU(3)_{\scriptscriptstyle L}\times SU(3)_{\scriptscriptstyle R}$ chiral symmetry of light quarks, see, e.g., \cite{tmyhyc1992,pc11992,ysdsmh2016} and references therein.  It is worth mentioning that pion interactions can be formulated in the pseudoscalar and axial forms that are equivalent in the nonrelativistic limit. Connection between the respective coupling constants for  the pion-nucleon interaction is provided by the classical Goldberger-Treiman relationship

\beq \label{glodbterim}
g_{\scriptscriptstyle\pi NN}=g^A_{\scriptscriptstyle NN}\frac{M_N}{F_\pi},
\eeq

\noindent
where $g^{\scriptscriptstyle A}_{\scriptscriptstyle NN}$ is the nucleon axial charge and $g_{\scriptscriptstyle\pi NN}$ is the pseudoscalar interaction constant.

Relationships of this type exist not only for diagonal interactions but also for nondiagonal vertices, for example, for the $\pi\Sigma_c\Lambda_c$ interaction. Axial form of the interaction is  dictated by the goldstone nature of pions and the axial charge can be calculated, at least in principle, see, e.g., \cite{dorgeb2001,hpkmw2017}.

Experimental data on the decay widths $\Sigma^{++}_c\to\Lambda_c\pi^+$ and $\Sigma_c^{0}\to \Lambda_c\pi^-$ \cite{pdg2018}, provides direct access to the interaction constant $g_{\scriptscriptstyle\pi\Sigma_c\Lambda_c}$. With the Lagrangian in Table~\ref{pionlgr} one obtains

\beq
\Gamma(\Sigma_c\to \Lambda_c+\pi)=\frac{g^2_{\scriptscriptstyle\pi\Sigma_c\Lambda_c}}{4\pi}\frac{k (E_{\scriptscriptstyle\Lambda_c}-M_{\scriptscriptstyle\Lambda_c})}{M_{\scriptscriptstyle\Sigma_c}},
\eeq

\noindent
where $k$ is the decay momentum and $E_{\Lambda_c}$ is the energy of the final $\Lambda_c$. We obtain $g_{\scriptscriptstyle\pi\Sigma_c\Lambda_c}=19.3$  from the decay $\Sigma^{++}_c\to\Lambda_c\pi^+$
and $g_{\scriptscriptstyle\pi\Sigma_c\Lambda_c}=19.1$  from the decay $\Sigma^{0}_c\to\Lambda_c\pi^-$. We used  the average $g_{\scriptscriptstyle\pi\Sigma_c\Lambda_c}=19.2$ (compare \cite{tmyhyc1992,lo2012,lsgz2017}) in the calculations above.

There is no experimental data for the $\Sigma_c\Sigma_c\pi$ coupling, so we have chosen a roundabout way to determine the respective interaction constant. As mentioned above axial interaction constants can be in principle calculated theoretically if one knows form factors of the respective axial currents. Unfortunately, currently there is no effective way to calculate these form factors in QCD\footnote{It could be a good problem for the lattice gauge theory calculations.}.  It was suggested long time ago \cite{dorgeb2001} to use the naive constituent quark model to calculate diagonal and transitional axial charges. The quark model predicts $g^a_{\scriptscriptstyle\Sigma_c\Lambda_c}=2/\sqrt{3}\sim 1.154$ to be compared with the value we calculate from the experimental decay widths  $g^a_{\scriptscriptstyle\Sigma_c\Lambda_c}=g_{\scriptscriptstyle\pi\Sigma_c\Lambda_c}F_\pi/M_{\scriptscriptstyle\Sigma_c}\sim 0.727$ (we neglect here mass difference of $\Sigma_c$ and $\Lambda_c$). It is clear that the accuracy of the quark model leaves much to be desired. We expect that it predicts ratios of axial constants more accurately than the axial constants themselves. The ratio of the axial constants $g^a_{\scriptscriptstyle\Sigma_c\Lambda_c\pi}$ and  $g^a_{\Sigma_c\Sigma_c}$ in the quark model is $g^a_{\scriptscriptstyle\Sigma_c\Lambda_c}/g^a_{\scriptscriptstyle\Sigma_c\Sigma_c}=1/\sqrt{3}$. The ratio of the respective pseudoscalar constants is proportional to the ratio of the axial constants and we obtain

\beq
g_{\scriptscriptstyle\pi\Sigma_c\Sigma_c}=\frac{1}{\sqrt{3}}g_{\scriptscriptstyle\pi\Sigma_c\Lambda_c}\approx 11.0.
\eeq

\noindent
We used this value in calculations  of the pentaquark decay widths. Other estimates of this constant $g_{\pi\Sigma_c\Sigma_c}\approx 10.76$ \cite{gz2015,lsgz2017}  are based on the assumption that $g_{\scriptscriptstyle\pi\Sigma_c\Sigma_c}=g_{\scriptscriptstyle\pi\Sigma\Sigma}$. This value is consistent with our estimate.

The axial interaction Lagrangian $\Sigma_c^*\Sigma_c\pi$ is in Table~\ref{pionlgr}. There is no $\gamma^5$ in this Lagrangian since contraction of the positive-parity Rarita-Schwinger spin-vector $\bar\Sigma^{*\mu}_c$, spinor $\Sigma^*_c$ and the axial vector $\partial_\mu\pi$ is a true scalar. The interaction has the gradient form, and the dimensionful interaction constant is proportional to the respective transitional axial constant. Naive quark model \cite{tmyhyc1992,lo2012} predicts that ratio of the $\Sigma^*_c\Sigma_c$ and $\Sigma_c\Sigma_c$  axial charges is $\sqrt{3}/2$. We parameterize the dimensionful interaction constant $\tilde g_{\scriptscriptstyle\pi\Sigma_c\Sigma_c^*}$ in terms of the dimensionless $g_{\scriptscriptstyle\pi\Sigma_c\Sigma_c^*}$

\beq
\tilde g_{\scriptscriptstyle\pi\Sigma_c\Sigma_c^*}=\frac{g_{\scriptscriptstyle\pi\Sigma_c\Sigma_c^*}}
{\sqrt{2M_{\scriptscriptstyle\Sigma_c^*}}\sqrt{2M_{\scriptscriptstyle\Sigma_c}}},
\eeq

\noindent
and calculate its value

\beq
g_{\scriptscriptstyle\pi\Sigma_c\Sigma_c^*}=\frac{\sqrt{3}}{2}\sqrt{\frac{M_{\scriptscriptstyle\Sigma_c^*}}
{M_{\scriptscriptstyle\Sigma_c}}}g_{\scriptscriptstyle\pi\Sigma_c\Sigma_c}
\approx0.88g_{\scriptscriptstyle\pi\Sigma_c\Sigma_c}=9.7.
\eeq

\noindent
This constant was used in calculations of the pentaquark decay width.

The constant $g_{\scriptscriptstyle\pi DD^*}$ is extracted from the experimental data on $(D^{*+}(2010)\to D^0\pi^+$ and $(D^{*+}(2010)\to D^+\pi^0)$ decays \cite{pdg2018}. The decay width calculated with the Lagrangian in Table~\ref{pionlgr} is

\beq
\Gamma(D^{*+})_{tot}=\frac{g^2_{\scriptscriptstyle\pi D D^*}}{8\pi}\frac{k^3}{M_{\scriptscriptstyle D^*}^2}.
\eeq

\noindent
Combined with the experimental data this expression gives $g_{\scriptscriptstyle\pi D D^*}$ cited in Table~\ref{pionlgr}.

The constant $g_{\scriptscriptstyle\pi D^*D^*}$ can be obtained from $g_{\scriptscriptstyle\pi D D^*}$ using the heavy quark relationship (see, e.g., \cite{cd2016}) $g_{\scriptscriptstyle\pi D^*D^*}=g_{\scriptscriptstyle \pi D D^*}/\sqrt{M_{\scriptscriptstyle D} M_{\scriptscriptstyle D^*}}$.

\subsection{Nucleon Interactions\label{nuclintlacn}}

\begin{table}[h!]
\caption{\label{fermboslgr} Nucleon interactions}
\begin{ruledtabular}
\begin{tabular}
{lll}
Interacting particles&Interaction Lagrangian& Coupling Constant
\\
\hline
$\Lambda_c ND$ &$ig_{\scriptscriptstyle\Lambda_cND}\bar N\gamma^5\Lambda_cD+H.c.$ & $g_{\scriptscriptstyle\Lambda_cND}=4.5$
\\
$\Sigma_cND$&$-ig_{\scriptscriptstyle\Sigma_cND}\bar N\gamma^5\bm\tau\cdot\bm\Sigma_cD+H.c.$ & $g_{\scriptscriptstyle\Sigma_c ND}=0.9$
\\
$\Sigma^*_cND$&
$g_{\scriptscriptstyle\Sigma_c^*ND}\bar N_i\tau^a_{ik} \Sigma_{ca}^{*\mu}\partial_\mu D^\dagger_k+H.c.$
&$g_{\scriptscriptstyle\Sigma_c^*ND}=0.55$~\mbox{GeV}$^{-1}$
\end{tabular}
\end{ruledtabular}
\end{table}

\subsubsection{$\Lambda_cND$ Interaction and $\Lambda_c$ Semileptonic  Decays}

Nucleon-charmed baryon-$D$-meson interaction constants  were obtained in the literature from the $SU(4)$ invariant Lagrangians, see, e.g., \cite{wlcmkzwl2001,gz2015,lsgz2017}, and references therein. The QCD sum rules were also used  to obtain the value of $g_{\scriptscriptstyle \Lambda_cND}$ \cite{fsnmn1998,fsnmn1999,fodfsnmn2001}, and produced $g_{\scriptscriptstyle \Lambda_cND}=7.9\pm 0.9$, what is significantly smaller than the $SU(4)$ prediction $g_{\scriptscriptstyle \Lambda_cND}=-13.7$ \cite{wlcmkzwl2001}.

In view of such uncertainty we would like to go another route and connect the $D$-meson interaction constants with the experimental data on the weak semileptonic decay $\Lambda_c\to \Lambda +e^++\nu_e$. The idea is to determine the constant $g_{\scriptscriptstyle\Lambda_c\Lambda D}$ from the experimental data on this decay and then use the $SU(3)$ flavor symmetry to calculate $g_{\scriptscriptstyle\Lambda_c N D}$ in terms of $g_{\scriptscriptstyle\Lambda_c\Lambda D}$.

Our approach  to finding $g_{\scriptscriptstyle\Lambda_c\Lambda D}$ is similar to the Goldberger-Treiman derivation of the relationship between the pseudoscalar interaction constant $g_{\scriptscriptstyle\pi NN}$ and the nucleon axial charge in \eq{glodbterim}. The decay $\Lambda_c\to \Lambda +e^++\nu_e$ is  described by six form factors

\beq
\label{formflcld}
\begin{split}
\langle \Lambda|\bar s\gamma^\mu c|\Lambda_c\rangle=&\bar \Lambda(p+q)\left[\gamma^\mu f_1(q^2)
+i\sigma^{\mu\nu} q_\nu f_2(q^2)+q^\mu f_3(q^2)\right]\Lambda_c(p),\\
\langle \Lambda|\bar s\gamma^\mu \gamma^5c|\Lambda_c\rangle=&\bar \Lambda(p+q)\left[\gamma^\mu g_1(q^2)
+i\sigma^{\mu\nu}q_\nu g_2(q^2)+q^\mu g_3(q^2)\right]\gamma^5\Lambda_c(p).
\end{split}
\eeq

\noindent
The transferred momentum squared $q^2$ is an invariant mass of the lepton pair and is kinematically bounded, $\sqrt{q^2}\leq M_{\Lambda_c}-M_\Lambda<M_D$. The lepton masses can be safely neglected in
the theoretical description of  the $\Lambda_c\to \Lambda +e^++\nu_e$ decay. Then the form factors $f_3$ and $g_3$ do not enter the  decay amplitude  due to conservation of the lepton currents.

The form factors have poles in $q^2$ at the masses of mesons with the respective quantum numbers but they are outside the kinematically allowed region. Let us calculate lowest mass pseudoscalar charmed meson $D$ contribution to the form factor $g_3$. We choose the pseudoscalar form for the $\Lambda_c\Lambda D$ interaction

\beq
{\mathcal L}_{\scriptscriptstyle P}=ig_{\scriptscriptstyle\Lambda_c\Lambda D}\bar\Lambda \gamma^5\Lambda_cD,
\eeq

\noindent
and use the standard definition for the $D$-meson decay constant

\beq
\langle0|\bar s\gamma^\mu\gamma^5c|D(p)\rangle=-if_{\scriptscriptstyle D}p^\mu,
\eeq

\noindent
where $f_D\approx 212$ MeV \cite{pdg2018}.

We approximate the pseudoscalar form factor of a pointlike axial current by the pole contribution

\beq \label{ansatzpe}
g_3=\frac{f_{\scriptscriptstyle D}g_{\scriptscriptstyle \Lambda_c\Lambda D}}{M_{\scriptscriptstyle D}^2-q^2},
\eeq

\noindent
and we would like to determine the constant $g_{\scriptscriptstyle \Lambda_c\Lambda D}$ from the experimental data on the semileptonic decay $\Lambda_c\to \Lambda +e^++\nu_e$. However, as mentioned above this form factor $g_3$ does not contribute to the $\Lambda_c\to \Lambda +e^++\nu_e$ decay. To overcome this difficulty we consider the $c$-quark to be heavy enough to use the heavy quark approximation. According to the heavy quark theory only two of the six form factors describing a typical heavy-light transition  in \eq{formflcld} are independent (see, e.g., \cite{manoh}), and

\beq \label{hevqirel}
f_1=g_1,\qquad f_2=f_3=g_2=g_3,
\eeq

\noindent
Thus the form factors $g_2$ and $f_2$ coincide with the form factor $g_3$ in \eq{ansatzpe}. Numerous models for the form factors $f_1$, $f_2$, $g_1$, and $g_2$ were constructed in \cite{cleo2005,yllmqhdww2009,gutsche2016,faustovcalk2016,li2017,husroberts2017} and compared with the experimental data on the $\Lambda_c\to \Lambda +e^++\nu_e$ decay.  Parameterizations of the form factors in these works depend on many parameters, and the simple pole ansatz in \eq{ansatzpe} was never used. We considered the $q^2$-dependent form factors in  \cite{cleo2005,yllmqhdww2009,gutsche2016,faustovcalk2016,li2017,husroberts2017} as experimental data and used the HQET relationships in \eq{hevqirel} to fit them not far from the pole with the simple pole ansatz in \eq{ansatzpe}\footnote{Some of the papers  \cite{cleo2005,yllmqhdww2009,gutsche2016,faustovcalk2016,li2017,husroberts2017} where written before the branching ratio $\Gamma(\Lambda_c\to\Lambda e^+\nu)/\Gamma_{tot}$ changed from 2\% to 3.6\% \cite{pdg2018}. To account for this change we rescaled the old results by the square root of the new and old branching ratios.}. As a result of these fits we obtained approximate values of the coupling constant $g_{\scriptscriptstyle \Lambda_c\Lambda D}$.

The $SU(3)$ flavor symmetry of light quarks combined with the heavy quark theory provides a relationship between $g_{\scriptscriptstyle \Lambda_c\Lambda D}$ and $g_{\scriptscriptstyle \Lambda_c N D}$. Light quarks in $\Lambda_c$ are in the flavor antitriplet $\bar{\bm 3}$ state, while $\Lambda$ is a member of the flavor octet $\bm 8$, and the light quark in the current in \eq{formflcld} (as well as in the $\bar D$-meson) is in the fundamental  flavor representation $\bm 3$. Then matrix elements of the flavor triplet $j_\beta$ currents between different flavor octet states and $\Lambda_c$ are proportional to  the Clebsch-Gordon coefficients

\beq
\langle H,a|j_\beta|H_c,\alpha\rangle\sim C^{\bm 8\alpha}_{\bar {\bm 3}\alpha,\bm 3 \beta},
\eeq

\noindent
where  $a$ is an $SU(3)$ octet index, while $\alpha$ and $\beta$ are antitriplet and triplet indices, respectively.  We use this relationship and \eq{formflcld} to obtain

\beq \label{clbschlnd}
g_{\scriptscriptstyle \Lambda_c N D} =\sqrt{\frac{3}{2}}g_{\scriptscriptstyle \Lambda_c\Lambda D}.
\eeq

\noindent
Fitting the form factors in \cite{cleo2005,yllmqhdww2009,gutsche2016,faustovcalk2016,li2017,husroberts2017} with the pole ansatz and using \eq{clbschlnd} we obtained  $g_{\scriptscriptstyle\Lambda_c N D}$ in the interval $3.5-5.5$. These values are much smaller than $g_{\scriptscriptstyle\Lambda_c N D}=13.7$ \cite{wlcmkzwl2001} from the $SU(4)$ symmetry widely accepted in the literature. We think that \cite{wlcmkzwl2001} strongly overestimates $g_{\scriptscriptstyle\Lambda_c N D}$ and used $g_{\scriptscriptstyle\Lambda_c N D}=4.5$ in the calculations above. This is, of course, only a not too accurate estimate of this coupling constant.

\subsubsection{$\Sigma_cND$ Interaction and Quark Model}

We estimate the coupling constant $g_{\scriptscriptstyle\Sigma_c N D}$ using the constant $g_{\scriptscriptstyle \Lambda_c N D}$  from \eq{clbschlnd}. Unfortunately, there is no $SU(3)$ flavor relationship between $g_{\scriptscriptstyle\Sigma_c N D}$ and $g_{\scriptscriptstyle\Lambda_c N D}$  since  light quarks in $\Sigma_c$ and $\Lambda_c$ are in different flavor  representations ($\bm 6$ and $\bar{\bm 3}$, respectively). One can obtain such a relationship in the constituent quark model. We start with the  proton, $\Lambda_c$, $\Sigma_c$ and $D$ quark model wave functions. Quarks in a nucleon are in the antisymmetric color state and hence the remaining wave function is symmetric. It is a product of a symmetric coordinate wave function  $f_{\scriptscriptstyle N}(\bm r_1,\bm r_2,\bm r_3)$ and a symmetric spin-flavor function. The proton wave function with spin up has the form (we suppress the antisymmetric color factor)

\beq
\begin{split}
\Psi_p^{\scriptscriptstyle\uparrow} =
\frac{1}{3\sqrt{2}}\bigl[&2u_1^{\scriptscriptstyle\uparrow}u_2^{\scriptscriptstyle\uparrow} d_3^{\scriptscriptstyle\downarrow}+
2u_1^{\scriptscriptstyle\uparrow}d_2^{\scriptscriptstyle\downarrow} u_3^{\scriptscriptstyle\uparrow}
+2d_1^{\scriptscriptstyle\downarrow}u_2^{\scriptscriptstyle\uparrow} u_3^{\scriptscriptstyle\uparrow}
-u_1^{\scriptscriptstyle\downarrow}u_2^{\scriptscriptstyle\uparrow} d_3^{\scriptscriptstyle\uparrow}
-u_1^{\scriptscriptstyle\downarrow}d_2^{\scriptscriptstyle\uparrow} u_3^{\scriptscriptstyle\uparrow}
-u_1^{\scriptscriptstyle\uparrow}u_2^{\scriptscriptstyle\downarrow} d_3^{\scriptscriptstyle\uparrow}\\
&-d_1^{\scriptscriptstyle\uparrow}u_2^{\scriptscriptstyle\downarrow} u_3^{\scriptscriptstyle\uparrow}
-u_1^{\scriptscriptstyle\uparrow}d_2^{\scriptscriptstyle\uparrow} u_3^{\scriptscriptstyle\downarrow}
-d_1^{\scriptscriptstyle\uparrow}u_2^{\scriptscriptstyle\uparrow} u_3^{\scriptscriptstyle\downarrow}
\bigr]f_{\scriptscriptstyle N}(\bm r_1,\bm r_2,\bm r_3).
\end{split}
\eeq

\noindent
Respectively, the $\Lambda_c$ and $\Sigma_c^{\scriptscriptstyle ++}$ wave functions (again with spin up) are

\beq
\begin{split}
\Psi_{\scriptscriptstyle\Lambda_c}^{\scriptscriptstyle\uparrow}=&
\frac{1}{2}c_1^{\scriptscriptstyle\uparrow}
\bigl[u_2^{\scriptscriptstyle\uparrow}d_3^{\scriptscriptstyle\downarrow}
+d_2^{\scriptscriptstyle\downarrow}u_3^{\scriptscriptstyle\uparrow}
-u_2^{\scriptscriptstyle\downarrow}d_3^{\scriptscriptstyle\uparrow}
-d_2^{\scriptscriptstyle\uparrow}u_3^{\scriptscriptstyle\downarrow}\bigr]
f_{\scriptscriptstyle\Lambda_c}(\bm r_1,\bm r_2,\bm r_3),\\
\Psi_{\scriptscriptstyle\Sigma_c^{\scriptscriptstyle++}}^{\scriptscriptstyle\uparrow} =&
\frac{1}{\sqrt{6}}\bigl[2c_1^{\scriptscriptstyle\downarrow}
u_2^{\scriptscriptstyle\uparrow}u_3^{\scriptscriptstyle\uparrow}-
c_1^{\scriptscriptstyle\uparrow}u_2^{\scriptscriptstyle\downarrow}u_3^{\scriptscriptstyle\uparrow}
-c_1^{\scriptscriptstyle\uparrow}u_2^{\scriptscriptstyle\uparrow}u_3^{\scriptscriptstyle\downarrow}\bigr]
f_{\scriptscriptstyle\Sigma_c}(\bm r_1,\bm r_2,\bm r_3),
\end{split}
\eeq

\noindent
where the coordinate wave functions $f_{\scriptscriptstyle\Lambda_c}(\bm r_1,\bm r_2,\bm r_3)$ and $f_{\scriptscriptstyle\Sigma_c}(\bm r_1,\bm r_2,\bm r_3)$ are symmetric with respect to the permutation $\bm r_2\leftrightarrow\bm r_3$.  The $D^0$-meson wave function is

\beq
\Psi^0_{\scriptscriptstyle D}=\frac{1}{\sqrt{2}}\left[c^{\scriptscriptstyle\downarrow}_1\bar{u}^{\scriptscriptstyle\uparrow}_2
+c^{\scriptscriptstyle\uparrow}_1\bar{u}^{\scriptscriptstyle\downarrow}_2 \right]f_{\scriptscriptstyle D}(\bm r_1,\bm r_2).
\eeq

Transitions $\Lambda_c\to N+D$ and $\Sigma_c\to N+D$ in the quark model happen when a heavy $c$-quark emits a hard gluon that creates a light quark-antiquark pair. The heavy spectator $c$-quark picks up the light antiquark and forms $D$-meson, and the light quark joins the remaining two light quarks to form a nucleon. Emission of a hard gluon followed by the creation of a light quark-antiquark pair is effectively described by a flavor singlet operator $S$.  Hence, the coupling constants  $g_{\scriptscriptstyle\Sigma_c N D}$ and $g_{\scriptscriptstyle\Lambda_c N D}$ are proportional to the overlap integrals

\beq
g_{\scriptscriptstyle\Sigma_c N D}=\langle DN|S|\Lambda_c\rangle, \qquad g_{\scriptscriptstyle\Lambda_c N D}=\langle DN|S|\Sigma_c\rangle.
\eeq

\noindent
We assume that the coordinate wave functions $f_{\scriptscriptstyle\Lambda_c}(\bm r_1,\bm r_2,\bm r_3)$ and $f_{\scriptscriptstyle\Sigma_c}(\bm r_1,\bm r_2,\bm r_3)$ coincide. Then

\beq
g_{\scriptscriptstyle \Sigma_c N D }=\frac{1}{6}g,\qquad g_{\scriptscriptstyle \Lambda N D
}=\sqrt{\frac{3}{2}}g,
\eeq

\noindent
where $g$ is one and the same overlap integral of the coordinate wave functions.

Thus we obtain the quark model prediction

\beq \label{lamcsigmscr}
g_{\scriptscriptstyle \Sigma_c N D }=\frac{g_{\scriptscriptstyle \Lambda_c N D }}{3\sqrt{3}}.
\eeq

\noindent
Numerically, $g_{\scriptscriptstyle \Sigma_c N D }\approx 1.35$ what is again less than  $g_{\scriptscriptstyle \Sigma_c N D }=2.69$ used in the literature, see, e.g., \cite{gz2015}.

\subsubsection{$\Sigma^*_cND$ Interaction and Heavy Quark Theory}

We consider $c$-quark as a heavy quark and use the heavy quark theory to connect coupling constants of the $\Sigma_cND$ and $\Sigma^*_cND$ interactions. Due to the heavy quark spin symmetry heavy-light isodoublet mesons ($c\bar q$), namely the pseudoscalar $D$-meson with spin zero and the vector $D^*$-meson with spin one form a spin doublet. This doublet in the covariant notation can be written as a two-index matrix field

\beq \label{heavymsf}
H^{(v)}(x)=\frac{1+\slashed v}{2}\left[\slashed D^{*(v)}+iD^{(v)}\gamma^5\right],
\eeq

\noindent
where $v^\mu$ is the heavy quark four-velocity, and  $D^{(v)}$ and $D^{*(v)}_\mu$ ($v^\mu D^{*(v)}_\mu=0$) are pseudoscalar and transverse vector field, respectively. The first index of the two-index matrix field $H^{(v)}$ is the spinor index of the heavy $c$-quark and the second is spinor index of the light quark (for notation and more details see \cite{manoh}).  The field  $H^{(v)}(x)$  transforms bilinearly under the Lorentz transformations.

Spin of light quarks in the isotriplet heavy baryons $(cqq)$ is one and these baryons form a spin doublet with spins $1/2$ and $3/2$. This doublet is described by the heavy quark theory field

\beq \label{superbarf}
S^{(v)}_\mu=-\frac{1}{\sqrt{3}}(\gamma_\mu+v_\mu)\gamma^5\Sigma_c^{(v)}+\Sigma^{*(v)}_{c\mu},
\eeq

\noindent
where the $\Sigma_c^{(v)}$ and $\Sigma_{c\mu}^{*(v)}$ are spinor and Rarita-Schwinger fields, respectively. Both fields satisfy the heavy quark theory Dirac equations $\slashed v\Sigma_c^{(v)}=\Sigma_c^{(v)}$ and $\slashed v\Sigma_{c\mu}^{(v)}=\Sigma_{c\mu}^{(v)}$. The Rarita-Schwinger field satisfies also the standard additional conditions $v^\mu\Sigma_{c\mu}^{*(v)}=\gamma^\mu\Sigma_{c\mu}^{*(v)}=0$, that are necessary to reduce the number of independent components of the field describing the particle with spin $3/2$ to four. Easy to see that due to transversality of the field $\Sigma_{c\mu}^{*(v)}$ the spin-doublet field $S^{(v)}_\mu$ satisfies the condition $v^\mu S^{(v)}_\mu=0$.

The simplest  interaction Lagrangian preserving all symmetries of the strong interactions has the form

\beq \label{heavyintlagr}
\mathcal{L}_P=ig \bar{S}^{(v)}_\nu\sigma^{\mu\nu}\gamma^5H^{(v)}\partial_\nu N+H.c.,
\eeq

\noindent
where $N$ is the four-component nucleon field.

In the logic of the heavy quark theory interaction with light degrees of freedom should not change velocity of the heavy quark, and emission of a light nucleon with small but nonzero velocity should be considered as a first order correction to the heavy quark limit. This explains why the derivative in the interaction Lagrangian in \eq{heavyintlagr} is applied to the nucleon field, what makes the interaction vertex proportional to the nucleon velocity. The interaction Lagrangian in \eq{heavyintlagr} is therefore by construction a first order correction to the heavy quark limit and we avoid a hard task of calculating corrections on the background of  large zero order term contributions.

We are looking for a relationship between the $\Sigma_cND$ and $\Sigma^*_cND$ interaction constants so  the term with $D^*$ in \eq{heavyintlagr} can be omitted, and effectively

\beq
H^{(v)}\to \frac{1+\slashed v}{2}iD^{(v)}\gamma^5.
\eeq

\noindent
Then after substitution of the explicit expression for the field  $\bar S^{(v)}_\mu$ in \eq{heavyintlagr} one obtains

\beq
\mathcal{L}_P\to ig
\left[\frac{1}{\sqrt{3}}\bar\Sigma^{(v)}_c\gamma^5(\gamma_\mu+v_\mu)+\bar\Sigma^{*(v)}_{c\mu}\right]
\sigma^{\mu\nu}\frac{1-\slashed v}{2}iD^{(v)}\partial_\nu N +H.c.,
\eeq

\noindent
The heavy quark theory $\Sigma^*_cND$  interaction term turns into

\beq
\mathcal{L}_{\scriptscriptstyle\Sigma^*_cND}
=-ig\bar\Sigma^{*(v)}_{c\mu}\partial^\mu D N+H.c.,
\eeq

\noindent
In the transformations leading to this expression we used the conditions on the field $\Sigma^{*(v)}_{c\mu}$ below \eq{superbarf}, the explicit expression $\sigma^{\mu\nu}=i(\gamma^\mu\gamma^\nu-g^{\mu\nu})$, and allowed ourselves integration by parts.  Obviously this heavy quark theory interaction  coincides with the respective  effective  Lagrangian in Table~\ref{fermboslgr}, and, hence $g_{\scriptscriptstyle\Sigma^*_cND}=g$.

Similar calculations with the field $\bar\Sigma^{(v)}_c$ lead to the heavy quark theory $\Sigma_cND$ interaction term

\beq \label{sigmaheql}
\mathcal{L}_{\scriptscriptstyle\Sigma_cND}
=ig\sqrt{3}\bar\Sigma^{(v)}_c\gamma^5 v^\nu D^{(v)}\partial_\nu N+H.c.
\eeq

\noindent
As discussed above this Lagrangian is a first order correction to the heavy quark limit due to the explicit derivative of the light nucleon field. Hence, it is legitimate to let $v^\mu=(1,\bm 0)$ in all other terms. Then only the time derivative proportional to the light nucleon mass survives in the expression above, and the interaction term in \eq{sigmaheql} coincides with the respective phenomenological Lagrangian in Table~\ref{fermboslgr}, and we conclude that (recall that $g_{\scriptscriptstyle\Sigma^*_cND}=g$)

\beq
g_{\scriptscriptstyle\Sigma^*_cND}=\frac{g_{\scriptscriptstyle\Sigma_cND}}{\sqrt{3}M_{\scriptscriptstyle N}}.
\eeq

\noindent
We use $g_{\scriptscriptstyle\Sigma_cND}$ calculated above and obtain $g_{\scriptscriptstyle\Sigma^*_cND}=0.55$ GeV$^{-1}$. This value is much smaller than $g_{\scriptscriptstyle \Sigma^*_c N D }=6.5$ GeV$^{-1}$ cited in  \cite{lsgz2017}.   The authors of \cite{lsgz2017} made an assumption that $g_{\scriptscriptstyle \Sigma^*_c N D }=g_{\scriptscriptstyle \Sigma^* N K }$. Thus assumption can be justified  in the framework of the heavy quark symmetry if one considers both the $s$- and $c$-quarks as heavy quarks.  In its turn $g_{\scriptscriptstyle \Sigma^* N K }$ was calculated in \cite{yocmkkn2008,dhhkm2011} from $SU(3)$ flavor symmetry. The value of $g_{\scriptscriptstyle \Sigma^*_c N D }$ obtained above is only an estimate but we expect it to be more reliable than the one in \cite{lsgz2017} since simultaneous use of the $SU(3)$ flavor symmetry and heavy quark theory for $s$- and $c$-quarks hardly can be justified.

\subsection{Charmonium interactions\label{charminlagco}}

\begin{table}[h!]
\caption{\label{bosbosgr} Charmonium interactions}
\begin{ruledtabular}
\begin{tabular}
{lll}
Interacting particles&Interaction Lagrangian&Coupling Constant
\\
\hline
$J/\psi DD$ & $ -ig_{\scriptscriptstyle \Psi DD} \psi_\mu \left(\partial_\mu DD^\dagger-D\partial_\mu D^\dagger
\right)$ &  $g_{\scriptscriptstyle DDJ/\psi}=7.44$\footnote{Generalized vector dominance, see Appendix A.3 and \cite{clm2013}. }
\\
$\psi'DD$ & $ -ig_{\scriptscriptstyle \Psi' DD} \psi'_\mu \left(\partial_\mu DD^\dagger-D\partial_\mu D^\dagger
\right)$ &  $g_{\scriptscriptstyle DD\psi'}=12.51$\footnote{Generalized vector dominance, see Appendix A.3 and \cite{clm2013}.  }
\\
$J/\psi DD^*$& $- g_{\scriptscriptstyle J/\psi D^*D} \epsilon^{\mu\nu\alpha\beta}\partial_\mu\psi_\nu\left(D^{*\dagger}_\alpha\overset{\leftrightarrow}\partial_\beta D-D^\dagger\stackrel{\leftrightarrow}\partial_\beta\bar D^*_\alpha\right) $& $g_{\scriptscriptstyle J/\psi D^*D}=2.49$ GeV$^{-1}$\footnote{Generalized vector dominance and heavy quarks symmetry, see Appendix A.3 and \cite{clm2013}.   }
\\
$\psi'DD^*$&$- g_{\scriptscriptstyle \psi' D^*D} \epsilon^{\mu\nu\alpha\beta}\partial_\mu\psi'_\nu\left(D^{*\dagger}_\alpha\overset{\leftrightarrow}\partial_\beta D-D^\dagger\stackrel{\leftrightarrow}\partial_\beta\bar D^*_\alpha
\right) $ &$g_{\scriptscriptstyle \psi'DD^*}=3.52$ GeV$^{-1}$\footnote{Generalized vector dominance and heavy quarks symmetry, see Appendix A.3 and \cite{clm2013}.}

\end{tabular}
\end{ruledtabular}
\end{table}

Generalized vector dominance and/or QCD sum rules can be used to  calculate $J/\psi$ and $\psi'$ interaction constants with $D$ meson, see e.g., \cite{mebmcfsn2012} for a review. The basic assumption  of the generalized vector dominance is that photon interacts with $D$ via transitions into virtual vector mesons. Consider vector meson $V$ that is a bound state of $c\bar c$ quarks. The  zero component of the $c$-quark electric current $j^\mu_{(c)}=Q_c\bar c\gamma^\mu c$ ($Q_c$ is the $c$-quark charge) measures electric charge of the $c$-quark in $D$ meson. At zero momentum transfer $\langle D|j^0_{(c)}|D\rangle\sim Q_c$. On the other hand due to vector dominance the same matrix element is proportional to $g_{\scriptscriptstyle DVD}(1/M_{\scriptscriptstyle V}^2)Q_cf_{\scriptscriptstyle V}M_{\scriptscriptstyle V}$, where $M_{\scriptscriptstyle  V}$ is the vector meson mass and its decay constant $f_{\scriptscriptstyle V}$ is defined by the relationship $\langle 0|\bar c\gamma^\mu c|V\rangle=f_{\scriptscriptstyle V}M_{\scriptscriptstyle V}\epsilon^\mu$. Comparing these two expressions for the current matrix element we obtain  $g_{\scriptscriptstyle DDV}=M_{\scriptscriptstyle V}/f_{\scriptscriptstyle V}$. The vector meson decay constant $f_V$ is determined from the partial decay width

\beq \label{decvectmes}
\Gamma(V\to e^+e^-)=\frac{4\pi\alpha^2}{3}\frac{f_{\scriptscriptstyle V}^2Q_c^2}{M_{\scriptscriptstyle V}},
\eeq

\noindent
and

\beq
f_{\scriptscriptstyle V}=\frac{1}{2\alpha Q_c}\sqrt{\frac{3M_{\scriptscriptstyle V}\Gamma(V\to e^+e^-)}{\pi}}.
\eeq

\noindent
Experimentally $\Gamma(J/\psi\to e^+e^-)=5.55\pm 0.14\pm0.02$ keV and  $\Gamma(\psi'\to e^+e^-)=2.33\pm 0.04$ keV \cite{pdg2018}. Then $f_\psi\approx416.3$ MeV and $f_{\psi'}\approx294.68$ MeV \cite{cd2016}, and

\beq
g_{\scriptscriptstyle J/\psi DD}=\frac{M_\psi}{f_\psi}=7.44, \qquad
g_{\scriptscriptstyle \psi' DD}=\frac{M_{\psi'}}{f_{\psi'}}=12.51.
\eeq

\noindent
The dimensionful constants   $g_{\psi' DD^*}$ and $g_{J/\psi DD^*}$ are calculated from the heavy quark relationships (see, e.g., \cite{clm2013})

\beq
g_{\scriptscriptstyle J/\psi DD^*}=\frac{g_{\scriptscriptstyle J/\psi DD}}{M_{\scriptscriptstyle J/\psi}}\sqrt{\frac{M_{\scriptscriptstyle D^*}}{M_{\scriptscriptstyle D}}},\qquad
g_{\scriptscriptstyle \psi' DD^*}=\frac{g_{\scriptscriptstyle \psi' DD}}{M_{\psi'}}\sqrt{\frac{M_{\scriptscriptstyle D^*}}{M_{\scriptscriptstyle D}}}.
\eeq

\end{document}